\documentstyle[12pt,epsfig]{article}
\newcommand{\bm}[1]{\mbox{\boldmath $#1$}}
\newcommand{\be}{\begin{equation}}
\newcommand{\ee}{\end{equation}}
\newcommand{\bea}{\begin{eqnarray}}
\newcommand{\eea}{\end{eqnarray}}

\newcommand{\Lup}{\Lambda^\uparrow} 
 
\newcommand{\nd}{\noindent}
\newcommand{\NP}[1]{{\it Nucl.\ Phys.}\ {\bf #1}}
\newcommand{\ZP}[1]{{\it Z.\ Phys.}\ {\bf #1}}
\newcommand{\PL}[1]{{\it Phys.\ Lett.}\ {\bf #1}}
\newcommand{\PR}[1]{{\it Phys.\ Rev.}\ {\bf #1}}

\begin{document}
\begin{flushright} 
DFTT 16/2001 \\ 
INFNCA-TH0105 \\ 
IPPP/01/25 \\
DCPT/01/50 \\
\end{flushright} 
\vskip 1.5cm
\begin{center}
{\bf Weak interactions in polarized semi-inclusive DIS}\\
\vskip 0.8cm
{\sf M. Anselmino$^1$, M. Boglione$^2$, U. D'Alesio$^3$, F. Murgia$^3$}
\vskip 0.5cm
{\it $^1$ Dipartimento di Fisica Teorica, Universit\`a di Torino and \\
          INFN, Sezione di Torino, Via P. Giuria 1, I-10125 Torino, Italy}\\
\vspace{0.3cm}
{\it $^2$ Department of Physics, University of Durham, Science Laboratories \\
          South Road, Durham DH11 3LE, United Kingdom}\\
\vspace{0.3cm}
{\it $^3$ INFN, Sezione di Cagliari and Dipartimento di Fisica,  
Universit\`a di Cagliari,\\
C.P. 170, I-09042 Monserrato (CA), Italy} \\
\end{center}

\vspace{1.5cm}

\begin{abstract}
We calculate, within pQCD parton model at leading orders, the 
expression of the polarization $P$ of spin 1/2 hadrons 
(typically $\Lambda$ baryons), produced in polarized semi-inclusive DIS
in all possible cases in which weak interactions are involved. 
We discuss how to gather new information on fragmentation and distribution 
functions and give numerical estimates in the cases for which data are or 
will soon be available.
\end{abstract}

\vspace{0.6cm}

~~~PACS numbers: 13.60.Hb, 13.85.Ni, 13.87.Fh, 13.88.+e
\newpage 
\pagestyle{plain}
\setcounter{page}{1}
\nd
{\bf 1. Introduction}
\vskip 6pt

The polarization of spin 1/2 baryons inclusively produced in polarized
Deep Inelastic Scattering processes may be useful, if measurable, 
to obtain new information on polarized distribution and fragmentation 
functions. A lot of attention has recently been dedicated to the 
self-revealing polarization of $\Lambda$'s and other hyperons 
\cite{noi1}-\cite{noi2}. Most papers, with the exception of Refs. 
\cite{kbv}, \cite{bjk} and \cite{mssy}, do not consider weak interaction 
contributions, due to lack of available experimental information.

NOMAD collaboration have recently published some very interesting results
\cite{nom} on the $\Lambda$ polarization in $\nu_\mu$ charged current 
interactions; more data might soon be available from high energy neutral
current processes at HERA, due to electro-weak interference effects.
It is then appropriate and timely to perform a systematical and comprehensive 
study of weak interaction contributions to the production and the polarization
of baryons in as many as possible DIS processes.
We stress that such contributions are an important source of new 
information, due to the natural neutrino polarization and to the selected 
couplings of $W$'s to pure helicity states. 
    
We consider weak interactions in the following processes:
\bea
\nu     \, p  &\to& \ell^-   \, \Lambda^{\uparrow} + X \quad 
{\rm (charged \; current)} \nonumber \\
\bar\nu \, p  &\to& \ell^+   \, \Lambda^{\uparrow} + X \quad 
{\rm (charged \; current)} \nonumber \\
\ell^- \, p   &\to&     \nu  \,\>\> \Lambda^{\uparrow} + X \quad
{\rm (charged \; current)} \nonumber \\
\ell^+ \, p   &\to& \bar\nu  \,\>\> \Lambda^{\uparrow} + X \quad
{\rm (charged \; current)} \nonumber \\
\nu     \, p  &\to& \nu      \,\>\> \Lambda^{\uparrow} + X \quad 
{\rm (neutral \; current)} \nonumber \\
\bar\nu \, p  &\to& \bar \nu \,\>\> \Lambda^{\uparrow} + X \quad 
{\rm (neutral \; current)} \nonumber \\
\ell \, p     &\to& \ell     \;\>\> \Lambda^{\uparrow} + X \quad 
{\rm (neutral \; current)} 
\nonumber 
\eea
where the lepton $\ell$ and the proton $p$ may or may not be polarized, 
whereas the neutrinos are obviously always polarized ($\lambda _\nu = -1/2$, 
$\lambda _{\bar \nu} = +1/2$).

In our calculations we take into account leading twist factorization theorem, 
Standard Model elementary interactions at lowest perturbative order and 
LO QCD evolution only. Consequently the cross-sections for the 
production of a hadron $B$ in the current fragmentation region
are given by
\be
\frac{d\sigma}{dx\,dy\,dz} = \sum_q q(x,Q^2) \, \frac{d\hat{\sigma}}{dy} \, 
D_{B/q}(z,Q^2)\,, \label{totcross}
\ee
where $q(x,Q^2)$ is the quark $q$ distribution function, $D_{B/q}(z,Q^2)$ 
is the fragmentation function of the quark into the detected hadron $B$, and 
$d\hat{\sigma}/dy$ is the elementary cross-section. The usual DIS variables
$x$, $y$ and $z$ are defined as $x = Q^2/2p\cdot q$,
$y=q \cdot p/\ell \cdot p$ and $z = p_B\cdot p/p \cdot q$ 
(see also Appendix B).

In Sections 2-4 we consider separately the different processes,
and derive explicit expressions for the polarization of a final 
baryon $B$ in terms of elementary dynamics, quark distribution and
fragmentation functions. In Section 5 we discuss how experimental
data could be used to obtain specific new information
and give predictions for several processes which might
be of interest in the near future.   
In Appendix A we give full information on the kinematical 
ranges and configurations for each of the experiments in progress or 
planned, used to derive our numerical results. 
In Appendix B we discuss mass effects in the fragmentation process,
to clarify differences and relationships between different definitions 
of the variable on which the fragmentation functions depend.    

\vskip 18pt
\goodbreak
\nd
{\bf 2. Charged current processes, \bm{\nu p \to \ell \Lup X}
and \bm{\ell p \to \nu \Lup X}}
\vskip 6pt

Let us consider first the neutrino initiated processes, 
$\nu p \to \ell \Lup X$; for them, there exist $4$ possible elementary 
contributions, corresponding to the interactions:
\bea
&\nu   \,    d_j& \!\!  \to \> \ell^- u_i       \,\nonumber \\
&\nu \, \bar u_i& \!\!  \to \> \ell^- \bar d_j  \,\nonumber \\
&\bar \nu \, u_i& \!\!  \to \> \ell^+ d_j       \,\nonumber \\
&\bar \nu \, \bar d_j& \!\! \to \> \ell^+ \bar u_i \,
\label{elem1}
\eea
where we use the notation
\be
u_i = u,c \nonumber  \quad\quad\quad d_j = d,s \>. \nonumber 
\ee

Neglecting quark masses one finds that there is only one non-zero helicity
amplitude $\hat M_{\lambda^{\,}_{\ell}, \lambda^{\,}_{q_i}; 
\lambda^{\,}_{\nu}, \lambda^{\,}_{q_j}}$ 
for each of the elementary processes in (\ref{elem1}), and precisely
\bea
&\hat M_{--;--} ^{\nu d_j \to \ell ^- u_i }& \,=\, 
\hat M_{++;++} ^{\bar \nu  \bar d_j \to \ell ^+ \bar u_i } \,=\, 
-\,\frac{4 \pi \alpha V_{ij}}{\sin^2\theta_W} \,\frac{1}{y + M_W^2/xs}\,,\\
&\hat M_{-+;-+} ^{\nu \bar u_i \to \ell ^- \bar d_j }& \,=\, 
\hat M_{+-;+-} ^{\bar \nu  u_i \to \ell ^+  d_j } \,=\, 
-\,\frac{4 \pi \alpha V_{ij}}{\sin ^2 \theta _W}\,\frac{1-y}{y + M_W^2/xs}
\,, 
\eea
where, according to usual SM rules,
\be
V_{ud} = V_{cs} = \cos\theta _C \quad\quad\quad
V_{us} = -V_{cd} = \sin\theta _C \,,\nonumber 
\ee
$\theta _W$ is the Weinberg angle, $\theta _C$ is the Cabibbo angle
and $V_{ij} = V_{ji}^*$. 

The elementary cross-sections are computed according to 
\be
\frac{d\hat \sigma_{\lambda \lambda ^{\prime}}}{dQ^2} = 
\frac{1}{16\pi x^2 s^2} \> \vert 
\hat M _{\lambda \lambda ^{\prime}; \lambda \lambda ^{\prime}} \vert ^2 = 
\frac{1}{sx} \> \frac{d\hat \sigma_{\lambda \lambda ^{\prime}}}{dy}\,,
\label{crosssection}
\ee
from which we obtain
\bea
\frac{d\hat \sigma_{--} ^{\nu d_j \to \ell ^- u_i}}{dy} = 
\frac{d\hat \sigma_{++} ^{\bar \nu  \bar d_j \to \ell ^+ \bar u_i }}{dy} &=& 
\frac{\pi \alpha ^2}{xs}\,\frac{\vert V_{ij}\vert ^2}{\sin ^4 \theta_W} 
\left( \frac{1}{y + M_W^2/xs} \right) ^2\,, \label{sig1} \\
\frac{d\hat \sigma_{-+} ^{\nu \bar u_i \to \ell ^- \bar d_j}}{dy} =
\frac{d\hat \sigma_{+-} ^{\bar \nu u_i \to \ell ^+ d_j }}{dy} &=& 
\frac{\pi \alpha ^2}{xs}\,\frac{\vert V_{ij}\vert ^2}{\sin ^4 \theta_W} 
\left( \frac{1-y}{y + M_W^2/xs} \right) ^2\,. \label{sig2}
\eea
Notice that both $\nu$ and $\bar \nu$ couple only to quarks with negative 
helicity and antiquarks with positive helicity. 

We can now compute the longitudinal polarizations $P_{[\nu,\ell]}$ and 
$P_{[\bar \nu,\ell]}$ for any spin $1/2$ baryon $B$ ($\Lambda$'s and 
$\bar \Lambda$'s for instance) produced in neutrino initiated, charged
current DIS scattering processes: 
\be
P_{[\nu,\ell]} (B) = 
\frac{d\sigma^{\nu p \to \ell ^- B_+ X} - 
      d\sigma^{\nu p \to \ell ^- B_- X}}
{d\sigma^{\nu p \to \ell ^- B_+ X} + 
      d\sigma^{\nu p \to \ell ^- B_- X}} \label{pnul}
\ee
and
\be
P_{[\bar \nu,\ell]} (B) = 
\frac{d\sigma^{\bar \nu p \to \ell ^+ B_+ X} - 
      d\sigma^{\bar \nu p \to \ell ^+ B_- X}}
{d\sigma^{\bar \nu p \to \ell ^+ B_+ X} + 
      d\sigma^{\bar \nu p \to \ell ^+ B_- X}} \>, \label{pbnul}
\ee
where $B_{\pm}$ denotes a baryon $B$ with helicity $\pm$.
 
In the most general case, when also the proton $p$ is polarized -- and we
denote by an apex $S$ its spin state -- from Eqs. (\ref{totcross}), 
(\ref{pnul}) and (\ref{pbnul}) we obtain:
\be
P^{(S)}_{[\nu,\ell]} (B) = 
-\frac{\sum_{i,j} [ (d_j)^{(S)}_- d\hat\sigma_{--}^{d_j \to u_i} 
\Delta D_{B/u_i} - (\bar u_i)^{(S)}_+ d\hat\sigma_{-+}^{\bar u_i \to \bar d_j}
\Delta D_{B/\bar d_j}]}
{\sum_{i,j} [ (d_j)^{(S)}_- d\hat\sigma_{--}^{d_j\to u_i} D_{B/u_i} +
(\bar u_i)^{(S)}_+ d\hat\sigma_{-+}^{\bar u_i \to \bar d_j} D_{B/\bar d_j}]}
\label{Pnu}
\ee
and 
\be
P^{(S)}_{[\bar \nu,\ell]} (B) =
-\frac{\sum_{i,j} [ (u_i)^{(S)}_- d\hat\sigma_{+-}^{u_i \to d_j} 
\Delta D_{B/d_j} -
(\bar d_j)^{(S)}_+ d\hat\sigma_{++}^{\bar d_j \to \bar u_i} 
\Delta D_{B/\bar u_i}]}
{[ (u_i)^{(S)}_- d\hat\sigma_{+-}^{u_i \to d_j} D_{B/d_j} +
(\bar d_j)^{(S)}_+ d\hat\sigma_{++}^{\bar d_j \to \bar u_i} D_{B/\bar u_i}]}
\,,
\label{Pantinu}
\ee
where an expression like $(q)^{(S)}_{\pm}$ stands for the number density
(distribution function) of quarks $q$ with helicity $\pm$ inside a proton 
with spin $S$, whereas $q_{\pm}$ alone refers, as usual, to a proton with 
+ helicity. The polarized fragmentation functions are defined as
\be
\Delta D_{B /q} \equiv D_{B_+/q_+} - D_{B_-/q_+} = D_{B_-/q_-} - D_{B_+/q_-} 
\,. 
\ee

If we now explicitely perform the sum over flavours in the numerators and 
denominators of Eqs. (\ref{Pnu}) and (\ref{Pantinu}), neglecting $c$ quark 
contributions, and use Eqs. (\ref{sig1}) and (\ref{sig2}), we obtain 
for longitudinally ($\pm$ helicity) polarized protons
\be
P_{[\nu,\ell]} ^{(\pm)} (B;x,y,z) = 
-\frac{[d_\mp + R\,s_\mp]\, \Delta D_{B/u} - 
(1-y)^2 \,\bar u_\pm \,[\Delta D_{B/\bar d} + R\, \Delta D_{B/\bar s}]}
{[d_\mp + R\,s_\mp] \, D_{B/u} +
(1-y)^2 \,\bar u_\pm \, [D_{B/\bar d} + R\, D_{B/\bar s}]} 
\label{Pnu+-}
\ee
and 
\be
P_{[\bar \nu,\ell]} ^{(\pm)} (B;x,y,z) = 
\frac{[\bar d_\pm + R\, \bar s_\pm]\, \Delta D_{B/\bar u} - 
(1-y)^2 \,u_\mp \, [\Delta D_{B/d} + R\, \Delta D_{B/ s}]}
{[\bar d_\pm + R\, \bar s_\pm] \, D_{B/\bar u} +
(1-y)^2 \, u_\mp \, [D_{B/ d} + R\, D_{B/s}]} \,,
\label{Pantinu+-}
\ee
where $R\, \equiv \sin ^2 \theta_C/\cos ^2 \theta_C \simeq 0.056$. 

In the simpler case in which the proton is unpolarized one replaces 
$q_+$ and $q_-$ with $q/2$ so that
%
Eqs. (\ref{Pnu+-}) and (\ref{Pantinu+-}) become respectively 
\be
P_{[\nu,\ell]} ^{(0)} (B;x,y,z) = 
-\frac{[d + R\,s] \, \Delta D_{B/u} - 
(1-y)^2 \, \bar u \, [\Delta D_{B/\bar d} + R\, \Delta D_{B/\bar s}]}
{[d + R\,s] \, D_{B/u} +
(1-y)^2 \, \bar u \, [D_{B/\bar d} + R\, D_{B/\bar s}]} 
\label{Pnu0}
\ee
and 
\be  
P_{[\bar \nu,\ell]} ^{(0)} (B;x,y,z) =  
 \frac{[\bar d + R\, \bar s] \, \Delta D_{B/\bar u} - 
(1-y)^2 \, u [\Delta D_{B/ d} + R\, \Delta D_{B/ s}]}
{[\bar d + R\, \bar s] \, D_{B/\bar u} +
(1-y)^2 \, u [D_{B/ d} + R\, D_{B/s}]} \,,
\label{Pantinu0}
\ee
in agreement with the results of Ref. \cite{mssy}.

The formulae given above hold for any baryon and antibaryon with spin $1/2$. 
If we specify the final hadron observed, further simplifications are possible.
Let's consider the case in which a $\Lambda$ baryon is produced, or,
in general, a baryon (rather than an antibaryon): in this case
we can neglect terms which contain both $\bar q$ distributions (in a proton)
and $\bar q$ fragmentations (into a $\Lambda$) as they are both 
small, in particular at large $x$ and $z$. Then we simply have:  
\be
P_{[\nu,\ell]} ^{(\pm)} (\Lambda;z) \, \simeq \, 
P_{[\nu,\ell]} ^{(0)} (\Lambda;z) \,\simeq \,  
-\frac{\Delta D _{\Lambda/u}}{D_{\Lambda/u}}\,, \label{PnuL}
\ee
\be
P_{[\bar \nu,\ell]} ^{(\pm)} (\Lambda;z) \, \simeq 
\, P_{[\bar \nu,\ell]} ^{(0)} (\Lambda;z) \,\simeq \,
- \frac{\Delta D _{\Lambda/d} + R\,\,\Delta D _{\Lambda/s}}{D_{\Lambda/d} 
+ R\,\, D _{\Lambda/s}}\,, \label{PantinuL}
\ee
and the polarizations, up to QCD evolution effects, become functions of 
the variable $z$ only, since any other term apart from the fragmentation 
functions cancels out. For Eq. (\ref{PantinuL}) to hold one should also avoid
large $y$ regions, due to the factor $(1-y)^2$ in Eq. (\ref{Pantinu0}). 

Eqs. (\ref{PnuL}) and (\ref{PantinuL}) relate the values of the 
longitudinal polarization $P(\Lambda)$ to a quantity with a clear physical 
meaning, {\it i.e.} the ratio $\Delta D_{\Lambda/q}/D_{\Lambda/q}$; 
this happens with weak charged current interactions -- while it cannot happen 
in purely electromagnetic DIS \cite{noi2} -- due to the selection of the quark 
helicity and flavour in the coupling with neutrinos. A measurement of 
$P(\Lambda)$ offers new direct information on the fragmentation process. 
We shall discuss further this point in Section 5.   

Similar results hold for the $\ell p \to \nu \Lup X$ processes; 
the contributing elementary interactions are:
\bea
&\ell^-   \,    u_i& \!\! \to \> \nu \, d_j        \nonumber \\
&\ell^- \, \bar d_j& \!\! \to \> \nu \, \bar u_i     \nonumber \\
&\ell^+   \,    d_j& \!\! \to \> \bar\nu \, u_i      \nonumber \\
&\ell^+ \, \bar u_i& \!\! \to \> \bar\nu \, \bar d_j 
\eea
with the same cross-sections as those computed in Eqs. (\ref{sig1}) and 
(\ref{sig2}):
\be
\frac{d\hat \sigma_{--} ^{\ell^- u_i      \to \nu d_j}}{dy} = 
\frac{d\hat \sigma_{++} ^{\ell^+ \bar u_i \to \bar\nu \bar d_j}}{dy} = 
\frac{d\hat \sigma_{--} ^{\nu d_j \to \ell^- u_i}}{dy} =
\frac{d\hat \sigma_{++} ^{\bar \nu \bar d_j \to \ell^+ \bar u_i}}{dy}\,,
\ee
\be 
\frac{d\hat \sigma_{+-} ^{\ell^+ d_j      \to \bar\nu u_i}}{dy} = 
\frac{d\hat \sigma_{-+} ^{\ell^- \bar d_j \to \nu \bar u_i}}{dy} = 
\frac{d\hat \sigma_{+-} ^{\bar\nu u_i \to \ell^+ d_j}}{dy} = 
\frac{d\hat \sigma_{-+} ^{\nu \bar u_i \to \ell^- \bar d_j}}{dy} \>\cdot 
\ee

The analogue of Eqs. (\ref{Pnu+-}) and (\ref{Pantinu+-}) is now
\be
P_{[\ell,\nu]} ^{(\pm)} (B;x,y,z) = 
\frac{(1-y)^2 \, [\bar d_\pm + R\, \bar s_\pm]\, \Delta D_{B/\bar u} - 
u_\mp \,[\Delta D_{B/d} + R\, \Delta D_{B/s}]}
{(1-y)^2 \, [\bar d_\pm + R\, \bar s_\pm] \, D_{B/\bar u} +
u_\mp \, [D_{B/d} + R\, D_{B/s}]} 
\label{Pl-+-}
\ee
and 
\be
P_{[\ell,\bar \nu]} ^{(\pm)} (B;x,y,z) = 
- \frac{(1-y)^2 \, [d_\mp + R\, s_\mp]\, \Delta D_{B/u} - 
\bar u_\pm \,[\Delta D_{B/\bar d} + R\, \Delta D_{B/\bar s}]}
{(1-y)^2 \, [d_\mp + R\, s_\mp] \, D_{B/u} +
\bar u_\pm \, [D_{B/\bar d} + R\, D_{B/\bar s}]} 
\label{Pl++-}
\ee
and similarly for the analogue of Eqs. (\ref{Pnu0}) and (\ref{Pantinu0})
(one simply replaces in the above equations the quark and antiquark 
helicity distributions with the unpolarized ones). 

In the case in which one can neglect antiquark contributions (as for 
$\Lambda$'s) one has again, as in Eqs. (\ref{PnuL}) and (\ref{PantinuL}), 
\be
P_{[\ell,\nu]} ^{(\pm)} (\Lambda;z) \, \simeq \, 
P_{[\ell,\nu]} ^{(0)} (\Lambda;z) \,\simeq \,
- \frac{\Delta D _{\Lambda/d} + R\,\,\Delta D _{\Lambda/s}}{D_{\Lambda/d} 
+ R\,\, D _{\Lambda/s}}\,, \label{PnuL2}
\ee
\be
P_{[\ell, \bar \nu]} ^{(\pm)} (\Lambda;z) \, \simeq 
\, P_{[\ell, \bar \nu]} ^{(0)} (\Lambda;z) \,\simeq \,
-\frac{\Delta D _{\Lambda/u}}{D_{\Lambda/u}} \label{PantinuL2} \>\cdot
\ee

\vskip 18pt
\goodbreak
\nd
{\bf 3. Neutral current neutrino processes, \bm{\nu p \to \nu \Lup X}} 
\vskip 6pt

There are $4$ different kinds of elementary interactions contributing to 
these processes
\bea
&\nu  \,    q&       \to \>\> \nu       \, q   \nonumber \\
&\nu  \, \bar q&     \to \>\>  \nu      \, \bar q \nonumber \\
&\bar \nu \, q&      \to \>\> \bar \nu  \, q \nonumber \\
&\bar \nu \, \bar q& \to \>\> \bar \nu  \, \bar q 
\label{elem2}
\eea
where $q$ can be either $u_j$ or $d_j$.

There are $2$ non-zero independent helicity amplitudes for each process in 
(\ref{elem2}). 
These lead, through Eq. (\ref{crosssection}), to the following elementary 
cross-sections
\bea
\frac{d\hat \sigma_{-+} ^{\nu q \to \nu q}}{dy} &=& 
\frac{d\hat \sigma_{+-} ^{\bar \nu \bar q \to \bar \nu \bar q}}{dy} \>=\>
\frac{\pi \alpha ^2}{4xs}\,
\frac{(2\,e_q \, \sin ^2 \theta _W)^2}{\sin ^4 \theta_W \cos ^4 \theta_W} 
\left( \frac{1-y}{y + M_Z^2/xs} \right)^2  \,,
\nonumber \\
\frac{d\hat \sigma_{--} ^{\nu q \to \nu q}}{dy} &=& 
\frac{d\hat \sigma_{++} ^{\bar \nu \bar q \to \bar \nu \bar q}}{dy} \>=\>
\frac{\pi \alpha ^2}{4xs}\,
\frac{(1-2\,|e_q| \, \sin ^2 \theta _W)^2}{\sin ^4 \theta_W \cos ^4 \theta_W} 
\left( \frac{1}{y + M_Z^2/xs} \right)^2 \,,
\nonumber \\
\frac{d\hat \sigma_{-+} ^{\nu \bar q \to \nu \bar q}}{dy} &=&
\frac{d\hat \sigma_{+-} ^{\bar \nu  q \to \bar \nu  q}}{dy} \>=\>
\frac{\pi \alpha ^2}{4xs}\,
\frac{(1-2\,|e_q| \, \sin ^2 \theta _W)^2}{\sin ^4 \theta_W \cos ^4 \theta_W} 
\left( \frac{1-y}{y + M_Z^2/xs} \right) ^2  \,,
\nonumber \\
\frac{d\hat \sigma_{--} ^{\nu \bar q \to \nu \bar q}}{dy} &=&
\frac{d\hat \sigma_{++} ^{\bar \nu q \to \bar \nu q}}{dy} \>=\>
\frac{\pi \alpha ^2}{4xs}\,
\frac{(2\,e_q \, \sin ^2 \theta _W)^2}{\sin ^4 \theta_W \cos ^4 \theta_W} 
\left( \frac{1}{y + M_Z^2/xs} \right) ^2 \,,
\label{sig4}
\eea 
where $e_q$ is the quark charge in units of the proton charge.

In analogy to what we did in the previous paragraph, the 
longitudinal polarization of the produced baryon $B$ is defined as
\be
P_{[\nu ,\nu]}(B) = 
\frac{d\sigma^{\nu p \to \nu B_+ X} - 
      d\sigma^{\nu p \to \nu B_- X}}
{d\sigma^{\nu p \to \nu B_+ X} + 
      d\sigma^{\nu p \to \nu B_- X}} 
\ee
and 
\be
P_{[\bar \nu ,\bar \nu]}(B) =  
\frac{d\sigma^{\bar \nu p \to \bar \nu B_+ X} - 
      d\sigma^{\bar \nu p \to \bar \nu B_- X}}
{d\sigma^{\bar \nu p \to \bar \nu B_+ X} + 
      d\sigma^{\bar \nu p \to \bar \nu B_- X}} \>\cdot
\ee

For the numerator and denominator of $P_{[\nu, \nu]}(B)$ and 
$P_{[\bar \nu, \bar \nu]}(B)$ separately, one obtains, for a generic spin 
state $S$ of the proton:
\bea
N_{[\nu,\nu]}^{(S)} (B) \!\!&=&\!\! 
 \sum \,\! _{_j} \left\{ \left[ 
(u_j)_+^{(S)} \, (1-y)^2 \, 16\,C^2 -
(u_j)_-^{(S)} \, (1 - 4\,C)^2 \right] \Delta D_{B/u_j}  
\right . \nonumber \\ 
&& \mbox{} \hspace{0.3cm} + \left. \left[ 
(d_j)_+^{(S)} \, (1-y)^2 \, 4\,C^2 -
(d_j)_-^{(S)} \,(1 - 2\,C)^2 \right] \Delta D_{B/d_j} 
\right .\nonumber \\ 
&& \mbox{} \hspace{0.3cm} + \left. \left[ 
(\bar u_j)_+^{(S)} \,(1-y)^2 \, (1 - 4\,C)^2 -
(\bar u_j)_-^{(S)} \, 16\,C^2 \right] \Delta D_{B/\bar u_j} 
\right .\nonumber \\ 
&& \mbox{} \hspace{0.3cm} + \left.\left[ 
(\bar d_j)_+^{(S)} \,(1-y)^2 \, (1 - 2\,C)^2 -
(\bar d_j)_-^{(S)} \, 4\,C^2 \right] \Delta D_{B/\bar d_j} 
\right\}\,,
\eea
\bea
D_{[\nu ,\nu]}^{(S)} (B) \!\!&=&\!\!  
\sum \,\! _{_j} \left\{ \left[ 
(u_j)_+^{(S)} \,(1-y)^2 \, 16\,C^2 +
(u_j)_-^{(S)} \,(1 - 4\,C)^2 \right] D_{B/u_j}  
\right . \nonumber \\ 
&& \mbox{} \hspace{0.3cm} + \left. \left[ 
(d_j)_+^{(S)} \,(1-y)^2 \, 4\,C^2 +
(d_j)_-^{(S)} \,(1 - 2\,C)^2 \right] D_{B/d_j} 
\right .\nonumber \\ 
&& \mbox{} \hspace{0.3cm} + \left. \left[ 
(\bar u_j)_+^{(S)} \,(1-y)^2 \, (1 - 4\,C)^2 +
(\bar u_j)_-^{(S)} \, 16\,C^2 \right] D_{B/\bar u_j} 
\right .\nonumber \\ 
&& \mbox{} \hspace{0.3cm} + \left.\left[ 
(\bar d_j)_+^{(S)} \,(1-y)^2 \, (1 - 2\,C)^2 +
(\bar d_j)_-^{(S)} \, 4\,C^2 \right] D_{B/\bar d_j} 
\right\}\,,
\eea
and 
\bea
N_{[\bar \nu ,\bar \nu]}^{(S)} (B) \!\!&=&\!\!  
 \sum \,\! _{_j} \left\{ \left[ 
(u_j)_+^{(S)} \, 16\,C^2 -
(u_j)_-^{(S)} \,(1-y)^2\,(1 - 4\,C)^2 \right] \Delta D_{B/u_j} 
\right . \nonumber \\ 
&&\mbox{} \hspace{0.3cm}+\left. \left[ 
(d_j)_+^{(S)} \, 4\,C^2 -
(d_j)_-^{(S)} \,(1-y)^2\,(1 -2\,C)^2 \right] \Delta D_{B/d_j} 
\right .\nonumber \\ 
&& \mbox{} \hspace{0.3cm}+ \left. \left[ 
(\bar u_j)_+^{(S)} \, (1 - 4\,C)^2 -
(\bar u_j)_-^{(S)} \,(1-y)^2 \, 16\,C^2 \right] 
\Delta D_{B/\bar u_j} \right .\nonumber \\ 
&&\mbox{} \hspace{0.3cm}+ \left.\left[ 
(\bar d_j)_+^{(S)} \, (1 - 2\,C)^2 -
(\bar d_j)_-^{(S)} \, (1-y)^2\, 4\,C^2 \right] 
\Delta D_{B/\bar d_j} \right\}\,,
\eea
\bea
D_{[\bar \nu ,\bar \nu]}^{(S)} (B) \!\!&=&\!\!  
\sum \,\! _{_j} \left\{ \left[ 
(u_j)_+^{(S)} \, 16\,C^2 +
(u_j)_-^{(S)} \,(1-y)^2\, (1 - 4\,C)^2 \right] D_{B/u_j}  
\right . \nonumber \\ 
&& \mbox{} \hspace{0.3cm}+ \left. \left[ 
(d_j)_+^{(S)} \, 4\,C^2 +
(d_j)_-^{(S)} \,(1-y)^2\, (1 - 2\,C)^2 \right] D_{B/d_j} 
\right .\nonumber \\ 
&& \mbox{} \hspace{0.3cm}+ \left. \left[ 
(\bar u_j)_+^{(S)} \,(1 - 4\,C)^2 +
(\bar u_j)_-^{(S)} \,(1-y)^2 \, 16\,C^2 \right] 
D_{B/\bar u_j} \right .\nonumber \\ 
&& \mbox{} \hspace{0.3cm}+ \left.\left[ 
(\bar d_j)_+^{(S)} \, (1 - 2\,C)^2 +
(\bar d_j)_-^{(S)} \, (1-y)^2 \, 4\,C^2 \right] D_{B/\bar d_j} 
\right\}\,,
\eea
where $C \equiv \sin^2\theta_W/3$. 

In the case of $\Lambda$ (or any baryon, rather than antibaryon) production, 
a simple expression for its longitudinal polarization $P$ can be 
obtained by neglecting the antiquark contributions and the terms proportional 
to $\sin^4 \theta_W$. For longitudinally polarized protons in this 
approximation we have
\be
P_{[\nu ,\nu]}^{(\pm)} (\Lambda) \simeq  
- \frac
{\sum_j [(u_j)_\mp \, (1 - 8\,C) \Delta D _{\Lambda/u_j} +
(d_j)_\mp \, (1 - 4\,C) \Delta D _{\Lambda/d_j}]}
{\sum_j [(u_j)_\mp \, (1 - 8\,C) D _{\Lambda/u_j} +
(d_j)_\mp \, (1 - 4\,C) D _{\Lambda/d_j}]} \,,
\ee
whereas for unpolarized proton, where $q_\pm \to q/2$, one obtains
\be
P_{[\nu ,\nu]}^{(0)} (\Lambda) \simeq 
- \frac
{\sum_j [u_j \, (1 - 8\,C) \Delta D _{\Lambda/u_j} +
d_j \, (1 - 4\,C) \Delta D _{\Lambda/d_j}]}
{\sum_j [u_j \, (1 - 8\,C) D _{\Lambda/u_j} +
d_j \, (1 - 4\,C) D _{\Lambda/d_j}]} \>\cdot \label{Pnunu}
\ee
Similar formulae, avoiding the large $y$ region, hold for 
$P_{[\bar \nu ,\bar \nu]}^{(\pm)} (\Lambda)$
and for $P_{[\bar \nu ,\bar \nu]}^{(0)} (\Lambda)$.
\vskip 18pt
\goodbreak
\nd
{\bf 4. Neutral current lepton processes, \bm{\ell p \to \ell \Lup X}}
\vskip 6pt

The possible elementary scatterings contributing to this process are of the 
form
\be
\ell \, q \to \ell \, q
\label{elem3}
\ee
where $\ell$ can be either $\ell ^+$ or $\ell ^-$ and $q$ can be any quark or 
antiquark.

There are 4 non-zero independent helicity amplitudes corresponding to the 
process in (\ref{elem3}).
Notice that in this case we must take into account the contributions of both 
weak and electromagnetic interactions, and the amplitudes are given by the 
sum of the two corresponding terms. According to Eq. (\ref{crosssection}), 
the elementary cross-sections can be written as 
\bea
\frac{d\hat \sigma _{\pm\pm}^{\ell q \to \ell q}}{dy} &=& 
\frac{\pi \alpha ^2}{16 x s} 
\left( N_{\pm\pm}^{\ell q} \; \frac{1}{y + M_Z^2/xs} - \frac{8\,e_q}{y} 
\right) ^2 \,,\nonumber \\
\frac{d\hat \sigma _{\pm\mp}^{\ell q \to \ell q}}{dy} &=& 
\frac{\pi \alpha ^2}{16 x s} 
\left( N_{\pm\mp}^{\ell q} \; \frac{1}{y + M_Z^2/xs} - \frac{8\,e_q}{y}
\right) ^2 (1-y)^2 \,, \label{lqlq}
\eea
where again $\ell$ can be either $\ell ^+$ or $\ell ^-$, $q$ can be 
any quark or antiquark $q=u_j,d_j$ and $e_q$ is the quark charge.
For the coefficients $N^{\ell q}$ we have ($C = \sin^2\theta_W/3$)
\bea
N_{++}^{\ell ^- u_j} \!\!&=&\!\! N_{+-}^{\ell ^- \bar u_j} = 
N_{-+}^{\ell ^+ u_j}  =  N_{--}^{\ell ^+ \bar u_j} =
\frac{-16\,C}{1 - 3\,C} \simeq -1.60 \,,
\nonumber \\
N_{+-}^{\ell ^- u_j} \!\!&=&\!\! N_{++}^{\ell ^- \bar u_j} = 
N_{--}^{\ell ^+ u_j}  =  N_{-+}^{\ell ^+ \bar u_j} =
\frac{4\,(1 - 4\,C)}{1 - 3\,C} \simeq 3.60 \,,
\nonumber \\
N_{-+}^{\ell ^- u_j} \!\!&=&\!\! N_{--}^{\ell ^- \bar u_j} = 
N_{++}^{\ell ^+ u_j}  =  N_{+-}^{\ell ^+ \bar u_j} =
\frac{8\,(1 - 6\,C)}{3\,(1 - 3\,C)} \simeq 1.87 \,,
\nonumber \\
N_{--}^{\ell ^- u_j} \!\!&=&\!\! N_{-+}^{\ell ^- \bar u_j} = 
N_{+-}^{\ell ^+ u_j}  =  N_{++}^{\ell ^+ \bar u_j} =
\frac{2\,(6\,C - 1)(1 - 4\,C)}{3\,C\,(1 - 3\,C)} \simeq -4.19 \,,
\label{nuj}
\eea
and
\bea
N_{++}^{\ell ^- d_j} \!\!&=&\!\! N_{+-}^{\ell ^- \bar d_j} = 
N_{-+}^{\ell ^+ d_j}  =  N_{--}^{\ell ^+ \bar d_j} =
\frac{8\,C}{1 - 3\,C} \simeq 0.80 \,,
\nonumber \\
N_{+-}^{\ell ^- d_j} \!\!&=&\!\! N_{++}^{\ell ^- \bar d_j} = 
N_{--}^{\ell ^+ d_j}  =  N_{-+}^{\ell ^+ \bar d_j} =
\frac{4\,(2\,C - 1)}{1 - 3\,C} \simeq -4.40 \,,
\nonumber \\
N_{-+}^{\ell ^- d_j} \!\!&=&\!\! N_{--}^{\ell ^- \bar d_j} = 
N_{++}^{\ell ^+ d_j}  =  N_{+-}^{\ell ^+ \bar d_j} =
\frac{4\,(6\,C - 1)}{3\,(1 - 3\,C)} \simeq -0.93 \,,
\nonumber \\
N_{--}^{\ell ^- d_j} \!\!&=&\!\! N_{-+}^{\ell ^- \bar d_j} = 
N_{+-}^{\ell ^+ d_j}  =  N_{++}^{\ell ^+ \bar d_j} =
\frac{2\,(1 - 6\,C)(1 - 2\,C)}{3\,C\,(1 - 3\,C)} \simeq 5.12 \,.
\label{ndj}
\eea
We can now proceed to the calculation of the longitudinal
polarization $P$ of the observed spin $1/2$ baryon
\be
P_{[\ell,\ell]} (B) = 
\frac{d\sigma^{\ell p \to \ell B_+ X} - 
      d\sigma^{\ell p \to \ell B_- X}}
     {d\sigma^{\ell p \to \ell B_+ X} + 
      d\sigma^{\ell p \to \ell B_- X}}\,,
\ee
where $\ell$ can be either $\ell ^+$ or $\ell ^-$.

$P_{[\ell,\ell]} (B)$ can be evaluated for any lepton and proton spin 
configuration. When both the proton $p$ and the lepton $\ell$ are 
longitudinally polarized (in helicity states), the polarization $P$ becomes
\be
P_{[\ell,\ell]} ^{(\pm,\pm)} (B) = \frac{
\sum _q \left[ q_{\pm} \, d\hat\sigma _{\pm +} ^{\ell q \to \ell q} -
q_{\mp} \, d\hat\sigma _{\pm -} ^{\ell q \to \ell q} \right] \Delta D _{B/q}}
{\sum _q \left[ q_{\pm} \, d\hat\sigma _{\pm +} ^{\ell q \to \ell q} +
q_{\mp} \, d\hat\sigma _{\pm -} ^{\ell q \to \ell q} \right] D _{B/q}} \,,
\ee 
where again $\ell$ stands for either $\ell ^+$ or $\ell ^-$, and the sum 
runs over all quarks and antiquarks, $q = u,d,s,\bar u, \bar d, \bar s, ...$. 

For longitudinally polarized leptons but unpolarized protons 
($q_{\pm} \to q/2$) we have
\be
P_{[\ell,\ell]}^{(\pm,0)} = \frac{ \sum _q q \, \left[ 
d\hat\sigma _{\pm +} ^{\ell q \to \ell q} -
d\hat\sigma _{\pm -} ^{\ell q \to \ell q} \right] \Delta D _{B/q}}
{\sum _q q \,\left[ 
d\hat\sigma _{\pm +} ^{\ell q \to \ell q} +
d\hat\sigma _{\pm -} ^{\ell q \to \ell q} \right] D_{B/q}} \>, \label{Pllpm}
\ee
while for unpolarized leptons but longitudinally polarized protons we have
\be
P_{[\ell,\ell]} ^{(0,\pm)} (B) = \frac{ \sum _q \left[
q_{\pm} \, ( d\hat\sigma _{++} ^{\ell q \to \ell q} +  
            d\hat\sigma _{-+} ^{\ell q \to \ell q})   - 
q_{\mp} \, ( d\hat\sigma _{+-} ^{\ell q \to \ell q} +
            d\hat\sigma _{--} ^{\ell q \to \ell q})
\right] \Delta D _{B/q}}
{\sum _q \left[ 
q_{\pm} \, ( d\hat\sigma _{++} ^{\ell q \to \ell q} +  
            d\hat\sigma _{-+} ^{\ell q \to \ell q})   + 
q_{\mp} \, ( d\hat\sigma _{+-} ^{\ell q \to \ell q} +
            d\hat\sigma _{--} ^{\ell q \to \ell q})
\right] D _{B/q} } \>\cdot
\ee 

Finally, the most interesting case is when neither the proton nor the lepton 
are polarized: in this case the longitudinal polarization of baryon $B$ is 
non-zero only due to parity violating weak contributions. We obtain
\be
P_{[\ell,\ell]}^{(0,0)}(B) = \frac{ \sum _q  q \left[
d\hat\sigma _{++} ^{\ell q \to \ell q} + 
d\hat\sigma _{-+} ^{\ell q \to \ell q} -
d\hat\sigma _{+-} ^{\ell q \to \ell q} - 
d\hat\sigma _{--} ^{\ell q \to \ell q}
\right] \Delta D_{B/q} }
{4\,\sum _q  q \, d\hat\sigma ^{\ell q \to \ell q} \, D_{B/q}}\,, \label{Punp}
\ee
where $d\hat\sigma ^{\ell q \to \ell q}$ is the unpolarized 
$\ell q \to \ell q$ cross-section 
\be
4\,d\hat\sigma ^{\ell q \to \ell q} =  
d\hat\sigma _{++} ^{\ell q \to \ell q} + 
d\hat\sigma _{-+} ^{\ell q \to \ell q} +
d\hat\sigma _{+-} ^{\ell q \to \ell q} + 
d\hat\sigma _{--} ^{\ell q \to \ell q}\>.
\ee
This effect might be measurable 
at HERA and numerical estimates will be given in the next Section.   

\vskip 18pt
\goodbreak
\nd
{\bf 5. Numerical estimates}
\vskip 6pt

In the previous Sections we have obtained explicit expressions for the 
polarization of baryons produced in DIS scatterings involving
weak interactions; we now use these formulae to give predictions in the
case of $\Lambda$ and $\bar\Lambda$ production, considering typical
kinematical configurations of ongoing or planned experiments. When 
convenient, we integrate over the actual physical ranges of some variables;
these are collected in Table 1 of Appendix A.
Our results should give a good comprehensive description of what to 
expect in all present or future experiments, and can be adapted to cover 
all realistic situations, according to different kinematical cuts and 
configurations.  

The polarization values depend on the known Standard Model dynamics,
on the rather well known partonic distributions, both unpolarized
and polarized, and on the quark fragmentation functions, again 
both unpolarized and polarized. The latter are not so well known and a choice 
must be made in order to give numerical estimates or in order to be able 
to interpret the measured values in favour of a particular set.

Unpolarized $\Lambda$ fragmentation functions are determined by fitting 
$e^+e^- \to \Lambda + \bar \Lambda + X$ experimental data, which are sensitive 
only to singlet combinations, like $D_{\Lambda/q} + D_{\Lambda/\bar q}
\equiv D_{(\Lambda + \bar\Lambda)/q}$.   
It is impossible to separate the fragmentation functions relative 
to $\Lambda$'s from those for $\bar\Lambda$'s in a model independent way; 
also flavour separation is not possible without appropriate initial 
assumptions, for example about $SU(3)$ flavour symmetry.
Polarized $\Lambda$ fragmentation functions are obtained by fitting 
the scarce data on $\Lambda$ polarization at LEP, sensitive only to 
non-singlet combinations like 
$\Delta D_{\Lambda/q} - \Delta D_{\Lambda/\bar q} = 
\Delta D^{val}_{\Lambda/q}$. Also in this case flavour separation has
to rely on models.  

Three typical sets of fragmentation functions, denoted as scenarios 1, 2 
and 3, and derived from fits to $e^+e^-$ data, are given in Ref. \cite{vog}. 
The unpolarized fragmentation functions are taken to be $SU(3)$ symmetric:
\be
D_{(\Lambda + \bar\Lambda)/u} = 
D_{(\Lambda + \bar\Lambda)/d} = 
D_{(\Lambda + \bar\Lambda)/s} =
D_{(\Lambda + \bar\Lambda)/\bar u} = 
D_{(\Lambda + \bar\Lambda)/\bar d} = 
D_{(\Lambda + \bar\Lambda)/\bar s} \>, \label{ds}
\ee
and have been derived for the combined production of $\Lambda$ and 
$\bar\Lambda$ and not for each of them separately.
 
For the polarized fragmentation functions they assume, at the initial scale 
$\mu^2$:
\bea
\Delta D_{\Lambda/s}(z, \mu^2) &=& z^\alpha \, 
D_{(\Lambda + \bar\Lambda)/s}(z, \mu^2)\,, \nonumber \\
\Delta D_{\Lambda/u}(z, \mu^2) &=& \Delta D_{\Lambda/d}(z, \mu^2)
= N_u \, \Delta D_{\Lambda/s}(z, \mu^2) \>. \label{deltas}
\eea
The three scenarios differ for the relative contributions of the 
strange quark polarization to $\Lambda$ polarization: $N_u = 0$, $N_u = -0.2$
and $N_u = 1$ for scenarios 1, 2 and 3 respectively \cite{vog}.
The ``unfavoured'' polarized fragmentations
$\Delta D_{\Lambda/\bar u}, \> \Delta D_{\bar\Lambda/u}$, etc. are
assumed to be negligible at the initial scale $\mu^2$, and are only generated 
by QCD evolution; it is then possible, for the polarized fragmentation 
functions, to obtain separately the contributions to $\Lambda$ and 
$\bar\Lambda$.

We adopt the above set of fragmentation functions as they are the least 
dependent on models, they have the proper QCD evolution, and the three 
scenarios are well representative of possible spin dependences.
We are then equipped with unpolarized fragmentation functions into 
$\Lambda + \bar\Lambda$ and with separate polarized fragmentation 
functions into $\Lambda$ and $\bar\Lambda$; we wish to give predictions 
and estimates for the polarizations of $\Lambda$ and $\bar\Lambda$,
which are measured separately. We then define the following computable
quantities:
\be
P^*(\Lambda) \equiv \frac{d\sigma^{\Lambda_+} - d\sigma^{\Lambda_-}}
{d\sigma^{\Lambda + \bar\Lambda}} = 
\frac{P(\Lambda)}{1 + T} \,,\label{pstarl}
\ee
and
\be
P^*(\bar\Lambda) \equiv \frac{d\sigma^{\bar\Lambda_+} - 
d\sigma^{\bar\Lambda_-}}
{d\sigma^{\Lambda + \bar\Lambda}} = 
P(\bar\Lambda) \> \frac{T}{1 + T} \,,\label{pstarbarl}
\ee
where the notations should be obvious and
\be
T = \frac{d\sigma^{\bar\Lambda}}{d\sigma^{\Lambda}} \, \cdot \label{capr}
\ee

Eqs. (\ref{pstarl}) and (\ref{pstarbarl}) allow to compute the values of
$P(\Lambda)$ and $P(\bar\Lambda)$ provided one can compute or measure
the ratio $T$:
\be
P(\Lambda) = (1 + T) \> P^*(\Lambda)\,,
\quad\quad
P(\bar \Lambda) = \left( 1 + \frac 1T \right) \> P^*(\bar\Lambda) \,.
\label{ptops}
\ee
Notice that $P$ is always larger in magnitude than $P^*$.

The ratio $T$ cannot be computed with the fragmentation set of Ref. 
\cite{vog}; it requires the knowledge of separate unpolarized fragmentation 
functions for $\Lambda$ and $\bar\Lambda$ and it depends on the 
chosen set. 
 
In Figs. 1-10 we show some results for $P^*(\Lambda, \bar\Lambda)$ for several
processes, with different initial spin configurations, and different 
kinematical conditions, corresponding to typical experimental setups, shown 
in Table 1 of Appendix A. These may easily 
be changed, according to experimental situations. Details are given
in the figure captions. We use the unpolarized distribution functions of 
Refs. \cite{unpdf}, the related polarized distribution functions 
of Ref. \cite{poldf} (we have explicitely checked that our numerical results
depend very little on the available sets of parton densities) and   
the fragmentation functions of 
Ref. \cite{vog}, mainly with scenarios 2 and 3. 

In Figs. 11-12 we give estimates for $P(\Lambda)$ [rather than 
$P^*(\Lambda)$], in the same cases of some of the previous figures; we have 
computed the ratio $T$ either with the $SU(3)$ symmetric set of unpolarized 
fragmentation functions of Ref. \cite{blt} or with a set derived from 
Ref. \cite{ind}, by imposing $SU(3)$ symmetry.   

The figure captions contain all relevant information about the
various cases; we give here some general comments about our results.

\begin{itemize}
\item 
We present results using mainly the fragmentation functions of 
scenarios 2 and 3 of Ref.~\cite{vog}, neglecting scenario 1, in which only 
$s$ quarks contribute to $\Lambda$ polarization. 
In fact, $P(\Lambda)$ is always negligible in this case,
given the small content of $s$ quarks in the nucleon target and the
SU(3)-symmetric nature of the unpolarized fragmentation functions utilized.
This can be seen by inspecting Eqs. (\ref{PnuL}), (\ref{PantinuL}) and 
(\ref{PnuL2}), (\ref{PantinuL2}) for charged current interactions and 
Eq. (\ref{Pnunu}) for neutral currents.
However, it is interesting to notice that for unpolarized fragmentation 
functions allowing for a strong $SU(3)$ symmetry breaking, like those of
Ref. \cite{ind}, the situation can be different, and scenario 1 might 
give sizeable asymmetries. According to Ref. \cite{ind},
$D_{\Lambda/u} = D_{\Lambda/d} \ll D_{\Lambda/s}$ and 
this can well compensate for the small factor $R$ in Eq. (\ref{PantinuL}), so
that, also in scenario 1, $P_{[\bar\nu, \ell]}$ can be large.

\item
Figs. 1, 2 and 11 summarize some of the most interesting features
of $\Lambda$ polarization in charged current interactions.
The large $\langle x \rangle$ values involved in NOMAD experiment imply
$T=d\sigma^{\bar\Lambda}/d\sigma^{\Lambda} \ll 1$, so that, from 
Eq. (\ref{pstarl}),
$P^*(\Lambda)$ is similar to $P(\Lambda)$ (compare Figs. 2 and 11) and 
follows closely the simple behaviour suggested by Eqs. (\ref{PnuL}) and 
(\ref{PantinuL}). $P^*(\bar\Lambda)$, instead, is suppressed 
by the the small ratio $T$, see Eq. (\ref{pstarbarl}); the actual estimated
value of $P(\bar\Lambda)$ is shown in Fig. 11 and is much larger.
Notice that a comparison between Figs. 1 and 2, {\it i.e.} between
Eqs. (\ref{PnuL}) and (\ref{PantinuL}), might give information on the 
ratios $C_q \equiv \Delta D_{\Lambda/q}/D_{\Lambda/q}$; for example,
the same value of $C_q$ for all flavours would result in 
$P_{[\nu,\ell]}(\Lambda) = P_{[\bar\nu,\ell]}(\Lambda)$.
On the other hand, largely different values of 
$P_{[\nu,\ell]}$ and $P_{[\bar\nu,\ell]}$ would certainly indicate a strong 
$SU(3)$ symmetry breaking in the fragmentation functions, with $s$ quark 
contributions dominating in order to compensate for the small $R$ factor 
in Eq. (\ref{PantinuL}).

Some data on $P^{(0)}_{[\nu,\ell]}(\Lambda, \bar\Lambda)$ are available from 
NOMAD collaboration \cite{nom}, but the errors and uncertainties are still too 
large to allow significant comparisons and to discriminate between different 
sets of fragmentation functions. 

\item
Fig. 3 gives values of $P^*(\Lambda)$ in kinematical regions dominated
by small $x$ values, so that one expects $T \simeq 1$ and  
$P^*(\Lambda,\bar\Lambda)\simeq P(\Lambda,\bar\Lambda)/2$, 
Eqs. (\ref{pstarl}) and (\ref{pstarbarl}).
The opposite signs of $P(\Lambda)$ and $P(\bar\Lambda)$ can be easily 
understood by looking at Eq. (\ref{Pl++-}) (with $q_{\pm}\to q/2$) 
and noticing that fragmentation into a baryon or an 
antibaryon favours the first or the second term in the numerator. 

\item
For neutrino charged currents we give numerical estimates only in the case 
of an unpolarized target. In fact, present intensities of neutrino beams 
require very large targets to reach reasonable luminosities and statistics, 
and this makes unpractical to polarize them. There are however proposals for 
neutrino factories with large intensities which will allow to consider the 
option of polarized targets \cite{rep}. 

\item
The $\Lambda$ polarizations for neutral currents shown in Figs. 4, 5, 6 and 7 
exhibit a similar behaviour for the four different kinematical setups 
considered. The differences are related to the different kinematical cuts and 
again to the value of the factor $T$. In particular, since $T \simeq 1$ for 
E665 and HERA kinematics, the $P^*$ are suppressed by a factor $\simeq 2$
with respect to the case of HERMES and COMPASS kinematics. Notice also that 
in these two cases there are sizeable variations depending on the different 
polarization states of the target. 

\item
The results presented in Figs. 8 and 12 show polarizations as functions
of $x$ (integrated over $z$) rather than $z$ (integrated over $x$): 
these test the dynamics of the partonic 
process and in particular the contribution of electro-weak interferences,  
in a neat and unusual way. The differences between positively and 
negatively charged leptons are entirely due to electro-weak effects; 
this is well visible at large $x$ (implying large $Q^2$),
where the curves for $e^+$ and $e^-$ differ sizeably.  
Moreover, from Eqs. (\ref{lqlq})-(\ref{ndj}) 
it is possible to evaluate analytically the zeros of the cross-section 
differences $d\hat\sigma_{\pm+}-d\hat\sigma_{\pm-}$
appearing in the numerator of Eq. (\ref{Pllpm}); one can show that real 
zeros for $0 < x < 1$ occur only for electromagnetic + weak contributions
and for positron beams.
The effective position of the zeros depends on $y$ (or alternatively on
$Q^2$) and for the dominating small $y$ values is around $x \simeq 0.04$ - 
$0.08$. Although the statistical errors increase sizeably for large $x$ 
values at HERA, the different behaviour shown at small and large $x$ values 
for positron and electron beams might probably be tested.

Fig. 12 shows the same plots as in Fig. 8, for the actual polarization
$P(\Lambda)$, estimated according to the comments in the figure caption, 
rather than for $P^*(\Lambda)$; it is interesting to note how the differences
between $P$ and $P^*$ vary with $x$, according to the observations we
have already made.

\item
Fig. 9 shows the parity violating longitudinal polarization
of $\Lambda$'s produced from unpolarized initial electrons and nucleons in 
NC processes; being a purely electro-weak effect it is more sizeable at 
very large $Q^2$ values, which are, however, accessible at HERA.
Also Fig. 10 shows some effects of electro-weak interferences, resulting 
in differences between plots of $P^{*\,(+,0)}_{[e,e]}$ for positrons and 
electrons.   
 
\end{itemize}

We have given a comprehensive discussion -- both theoretical (at LO) and 
phenomenological -- of the polarization of $\Lambda$'s and $\bar\Lambda$'s
produced in the current fragmentation region of DIS processes, both with 
neutral and charged currents. Our results can be exploited to gather new
information about polarized fragmentation functions, to improve our knowledge 
about polarized parton densities \cite{spfl}
and to test fundamental features of
electro-weak elementary interactions. Several experiments are either running 
or being planned, which will precisely look at these semi-inclusive DIS
processes; our study should help in the analysis of the forthcoming data.   

\vskip 18pt
\goodbreak
\nd
{\bf Acknowledgements}
\vskip 6pt
We would like to thank J.T. Londergan and A.W. Thomas for several
discussions; M.A. is grateful to the Special Research Centre for the 
Subatomic Matter of Adelaide (Australia) for hospitality and support 
during a period in which this paper was in preparation. 
M.B. is most grateful for partial support from the EU-TMR Program, Contract 
No. CT98-0169 and wishes to thank the Dept. of Theoretical Physics of 
Torino University for hospitality and travel support. 
U.D. and F.M. thank COFINANZIAMENTO MURST-PRIN for partial support.

\newpage

\nd
{\bf Appendix A -- Experimental setups and kinematical cuts}
\vskip 6pt

In our analysis we have considered most present and forthcoming 
experiments, which cover many different kinematical configurations.
For the reader's convenience, we collect and summarize here the 
corresponding experimental setups, with their kinematical ranges. 

The main variables which specify the various setups are listed 
below, while the kinematical values and cuts for the different 
experiments are given in Table 1.

\vspace{10pt}

\begin{tabular}{ll}
$E_l$:  & incoming lepton energy, in the Laboratory reference frame \\
$\sqrt{s}$:    & total energy, in the lepton-proton c.m. frame \\
$W$:	& total energy, in the virtual boson-proton c.m  frame \\
$E_{l'}$:& outgoing lepton energy, in the Laboratory frame \\
$\theta_{l'(h)}$:& outgoing lepton (hadron) scattering angle, in the 
Laboratory frame\\
$E_{h}$: & outgoing hadron energy, in the Laboratory frame\\
$p_T$: & transverse hadron momentum (w.r.t. the lepton direction)\\
$\eta$: & = $-$ ln$\!$ tan $\!(\theta_h/2)$, pseudorapidity,
        in the Laboratory frame\\
\end{tabular}

\vspace{6pt}

$x=Q^2/2q\cdot p$, $y= q\cdot p / \ell \cdot p $, $z=p_h\cdot p/q\cdot p$
are the usual invariant variables for semi-inclusive DIS hadron production.

\vspace{10pt}

\begin{center}
 \begin{tabular}{ccccccc}
  \hline\hline
   \noalign{\vspace{8pt}}
 	     & HERMES   & COMPASS    &   E665  &   NOMAD   &  HERA 
							   &  HERA*   \\
\noalign{\vspace{4pt}}\hline\noalign{\vspace{4pt}}

$E_l$ [GeV]  & 27.6     & 200        &   470   &   48.8    &  27.6
							   &  27.6    \\
\noalign{\vspace{4pt}}
\noalign{\vspace{4pt}}
$\sqrt{s}$ [GeV]& 7.26 & 19.4       &   29.7  &  9.6      &  300
							  &  300      \\
\noalign{\vspace{4pt}}
$x$	& 0.023-08 & $>$ 0.01   & $(10^{-3})$-0.1& 0.22   & $>$ 0.004
							  & $>$ 0.01  \\
\noalign{\vspace{4pt}}
$y$	    & $<$ 0.85 & 0.1-0.9    & 0.1-0.8 &   0.48    & 0.04-0.95 
							  & 0.1-0.95  \\
\noalign{\vspace{4pt}}
$z$	    & 0.2-0.7  & 0.2-0.9    & 0.1-0.95& 	  & $>$ 0.1   
							  & $>$ 0.1   \\
\noalign{\vspace{4pt}}
$Q^2$ [GeV$^2$]& 1-24  & $>$ 4	    & (1)-2.5 &   9       & 10-2000   
						          & 200-$10^4$\\
\noalign{\vspace{4pt}}
$W$ [GeV]   & $>$ 2    &            &	      &  5.8      &           
							  &	      \\
\noalign{\vspace{4pt}}
$E_{l'}$ [GeV]& $>$ 4.1&	    & $<$ 420 & 	  & $>$ 10    
							  & $>$ 10    \\
\noalign{\vspace{4pt}}
$\theta_{l'}$ [rad]& 0.04-0.22&     &	      &	          &           
							  &	      \\
\noalign{\vspace{4pt}}
$E_{h}$ [GeV]& $>$ 2   & $>$ 5	    & $>$ 4   &	          &           
							  &	      \\
\noalign{\vspace{4pt}}
$p_T$	[GeV]&         &	    &	      &	  	  & $>$ 0.5   
							  &           \\
$\eta$	     &         &	    &	      &		  & --1.5-1.5  
							  &           \\
\noalign{\vspace{8pt}}\hline\hline
\noalign{\vspace{12pt}}
 \end{tabular}
\end{center}
\nobreak
\begin{center}
{\bf Table 1:\ } Summary of the experiments and the corresponding 
kinematical setups.
\end{center}

\goodbreak

Wherever possible we have considered kinematical cuts 
identical to those already adopted or planned for the related experiments;
the $Q^2$ range for the E665 experiment at SLAC
($0.25 < Q^2 < 2.5$ GeV$^2$) reaches too low $Q^2$ values for our
leading order analysis, based on factorization theorem, and
we have adopted the range $1.0 < Q^2 < 2.5$ GeV$^2$
(this influences also the lower cut on $x$, of course).

Notice also that with HERA we mean both H1 and ZEUS typical setups 
at intermediate $Q^2$ values, while
with HERA* we refer to setups with very high $Q^2$ values,
as required for the study of electro-weak interference effects.

For NOMAD experiment, all kinematical variables are fixed to the
corresponding average value \cite{nau}.

\vskip 18pt
\nd
{\bf Appendix B -- Mass correction effects}
\vskip 6pt

In this paper fragmentation functions are always expressed as a function
of $z=p\cdot{p_h}/p\cdot{q}$, where $p$, $p_h$, $q$ are the four-momenta
of the target proton, the produced hadron, and the virtual boson respectively.
In the case of semi-inclusive DIS, at LO and in collinear configuration,
$z$ coincides with the light-cone momentum fraction of the parent
parton carried by the observed hadron, $\xi = p_{h}^+/p_q^+$. 
There are in general several other variables that can be considered;
depending on the specific process under study, they can be more
or less suitable than $z$ to, {\it e.g.}, describe the process from the
experimental point of view or to show scaling properties of 
observables, like cross-sections.
In this Appendix, we shortly review the definition of these
variables and give the connection among them.
It is important to notice that at very large energy scales $E$, when the
mass of the observed hadron $M_h$ can be safely neglected, all these variables
coincide (excluding the regions where they are comparable to $M_h/E$).
However, kinematics for most of the running or forthcoming
experiments on semi-inclusive hadron production, which is the main
subject of this paper, are such that mass corrections can be relevant.
We always neglect corrections due to the mass of the proton target,
even though they might have some effects in particular kinematical ranges.

Let us first briefly summarize the situation in the case of
$e^+e^-\to\Lambda\,X$ process, which is used to fix the set of
$\Lambda+\bar\Lambda$ fragmentation functions largely adopted in this
paper. The variables usually utilized are:
\bea
x_{_E} &=& \frac{2\,p_h\cdot{q}}{Q^2} = \frac{2E_h}{\sqrt{s}}\,,\nonumber\\
x_p &=& \frac{2\,|\mbox{\boldmath $p$}_h|}{Q} = 
        \frac{2\,|\mbox{\boldmath $p$}_h|}{\sqrt{s}} \,, \label{xexp} \\
\xi &=& \frac{p_h^+}{p_q^+} \,, \nonumber
\eea
\noindent
where $\mbox{\boldmath $p$}_h$ is the hadron three-momentum in the
$e^+e^-$ c.m. reference frame.
$x_{_E}$ and $x_p$ are usually adopted by the experimentalists, while
$\xi$ is more commonly used by the theorists.
At large energies, like in $e^+e^-$ collisions at the $Z_0$ pole,
and considering $x_{_E} > 0.1$, which is also required for other
theoretical reasons, see Ref.~\cite{vog}, mass effects are in fact negligible 
and all these variables can be safely assumed to be
equivalent; the fragmentation function dependence on $\xi$ can then
be directly identified with the $x_{_E}$ dependence shown by the experimental
results.
When mass corrections are relevant, the connection among the variables
defined in Eqs. (\ref{xexp}) is given, at leading order, as follows:
\bea
x_p &=& x_{_E}\,\beta \,, \nonumber\\
\xi &=& x_{_E}\,\frac{1+\beta}{2}\,,
\label{xpxi}
\eea
\noindent
where the factor $\beta$ is defined as
\be
\beta = \left(\,1-\frac{4M_h^2}{x_{_E}^2\,s}\,\right)^{1/2} \>\cdot
\label{beta}
\ee
Let us now consider the case of semi-inclusive DIS, in the virtual 
boson-target proton c.m. reference frame, for hadron production in the 
current fragmentation region ($x_F > 0$). Usual variables are:
\bea
z &=& \frac{p\cdot p_h}{p\cdot q}=
      \frac{E_h+|\mbox{\boldmath $p$}_h|}{W}\,,\nonumber\\
x_{_F} &=& \frac{2\,p_L}{W} = \frac{2\,|\mbox{\boldmath $p$}_h|}{W}\,,\\
z' &=& \frac{E_h}{E_{q'}} = \frac{E_h}{(1-x)E_p} =
       \frac{2E_h}{W}\,,\nonumber
\label{zxf}
\eea
\noindent
where $E_{q'}$ is the energy of the parent quark in the process
$q'\to h+X$.
$x_{_F}$ and $z$ are the variables usually adopted by the experimentalists.
However, as shown in Ref.~\cite{vog}, the appropriate
scaling variable for semi-inclusive DIS is $z'$ rather than
$x_{_F}$. 

Defining $\epsilon=M_h/W$, the ranges of variation of the three
variables are
\bea
x_{_F} &\in& \left[\,0,\,\left(1-4\epsilon^2\right)^{1/2}\,\right]\,,
\nonumber\\
z' &\in& \Bigl[\,2\epsilon, \,1\,\Bigr]\,,\\
z &\in& \left[\,\epsilon, \,\frac{1}{2}\,\Bigl\{\,1+
        \left(1-4\epsilon^2\right)^{1/2}\,\Bigr\}\,\right]\,.\nonumber
\label{range}
\eea

The expressions of the three variables as a function of the other two are
\bea
x_{_F}\!\!\! &=& \!\!\! 
z'\left(1-4\frac{\epsilon^2}{z'^2}\right)^{1/2} \quad\quad \quad\quad\quad 
\> \> x_{_F} = z\left(1-\frac{\epsilon^2}{z^2}\right) \,,\label{conv1} \\
z' \!\! \! &=& \!\!\! 
x_{_F}\left(1+4\frac{\epsilon^2}{x_{_F}^2}\right)^{1/2} \quad\quad\quad\quad
\quad \> \> z' = z\left(1+\frac{\epsilon^2}{z^2}\right) \,,\label{conv2} \\
z \!\!\! &=& \!\!\! 
x_{_F}\,\frac{1}{2}\left[1+\left(1+4\frac{\epsilon^2}{x_{_F}^2}
      \right)^{1/2}\right]\quad\quad 
z = z'\,\frac{1}{2}\left[1+\left(1-4\frac{\epsilon^2}{z'^2}
      \right)^{1/2}\right]\,. \label{conv3}
\eea

If we start from, {\it e.g.}, a cross-section evaluated in our formalism
(we omit here the dependence on $x$ and $y$)
\be
\frac{d\sigma}{dz}\propto D_h(z)\,,
\label{dsdh}
\ee
\noindent
the corresponding cross-section expressed as a function of $x_{_F}$
will be given by
\be
\frac{d\sigma}{dx_{_F}}=\frac{dz}{dx_{_F}}\,
\frac{d\sigma}{dz}\propto D_h[z(x_{_F})]\,,
\label{xftoz}
\ee
\noindent
where, apart from the overall rescaling factor $dz/dx_{_F}$,
one must keep into account
that fragmentation functions, obtained in the variable $z$, are to be
rescaled to the $z$ value corresponding to $x_{_F}$.

In variables like polarizations, given as ratios of two cross-sections,
the overall rescaling factors cancel out and the remaining effect is the
rescaling between the two variables in the fragmentation
functions, according to Eqs.~(\ref{conv1})-(\ref{conv3}).

Of course, if the average value of the polarization over a given
kinematical region is required, the appropriate overall
rescaling factors, like $dz/dx_{_F}$ in Eq.~(\ref{xftoz}),
have to be taken into account in the kinematical integrations.

\normalsize

\newpage

\noindent {\bf Figure captions}

\vspace{12pt}

\noindent{\bf Fig. 1:\ }
$P^{\ast\,(0)}_{[\nu,\mu]}$ for $\Lambda$ (solid lines) and
$\bar\Lambda$ (dashed lines) hyperons, as a function of $z$,
with a kinematical setup typical of NOMAD experiment at CERN
(see Table 1 for details).
Results are given for scenarios 2 and 3 of the polarized $\Lambda$ 
fragmentation functions (FF) of Ref.~\cite{vog}. Results
with scenario 1 are almost negligible. Unpolarized $(\Lambda+\bar\Lambda)$
FF are from Ref.~\cite{vog}; unpolarized GRV \cite{unpdf}
partonic distributions have been used for the proton target.
Since $\langle{x}\rangle$ is large for this kinematical
configuration, we expect
$T=d\sigma^{\bar\Lambda}/d\sigma^\Lambda \ll 1$ and, as a consequence,
$P^{(0)}_{[\nu,\mu]}(\Lambda)\simeq
P^{*\,(0)}_{[\nu,\mu]}(\Lambda)$, while
$P^{(0)}_{[\nu,\mu]}(\bar\Lambda) \gg
P^{*\,(0)}_{[\nu,\mu]}(\bar\Lambda)$ (see text for more details).

\vspace{8pt}

\noindent{\bf Fig. 2:\ }
$P^{*\,(0)}_{[\bar\nu,\mu]}$ for $\Lambda$ (solid lines) and
$\bar\Lambda$ (dashed lines) hyperons, as a function of $z$,
with a kinematical setup typical of NOMAD experiment at CERN
(see Table 1 for details).
Results are given for scenarios 2 and 3
of the polarized $\Lambda$ FF of Ref.~\cite{vog}. Results
with scenario 1 are almost negligible. Unpolarized $(\Lambda+\bar\Lambda)$
FF are from Ref.~\cite{vog}; unpolarized GRV \cite{unpdf}
partonic distributions have been used for the proton target.
Since $\langle{x}\rangle$ is large for this kinematical
configuration, we expect
$T=d\sigma^{\bar\Lambda}/d\sigma^\Lambda \ll 1$ and, as a consequence,
$P^{(0)}_{[\bar\nu,\mu]}(\Lambda)\simeq
P^{*\,(0)}_{[\bar\nu,\mu]}(\Lambda)$, while
$P^{(0)}_{[\bar\nu,\mu]}(\bar\Lambda) \gg
P^{*\,(0)}_{[\bar\nu,\mu]}(\bar\Lambda)$
(see text and Fig. 11 for more details).

\vspace{8pt}

\noindent{\bf Fig. 3:\ }
$P^{*\,(0)}_{[e,\bar\nu]}$ for $\Lambda$ (solid lines) and
$\bar\Lambda$ (dashed lines) hyperons, as a function of $z$,
with a kinematical setup typical of HERA experiments at DESY
(see Table 1 for details).
Results are given for scenarios 2 and 3
of the polarized $\Lambda$ FF of Ref.~\cite{vog}. Results
with scenario 1 are almost negligible. Unpolarized $(\Lambda+\bar\Lambda)$
FF are from Ref.~\cite{vog}; unpolarized GRV \cite{unpdf}
partonic distributions have been used for the proton target.
Since $\langle{x}\rangle$ is small for HERA kinematical
configurations, we expect
$T=d\sigma^{\bar\Lambda}/d\sigma^\Lambda \simeq 1$ and, as a consequence,
$P^{(0)}_{[e,\bar\nu]}(\Lambda,\bar\Lambda)\simeq
2\,P^{*\,(0)}_{[e,\bar\nu]}(\Lambda,\bar\Lambda)$
(see text for more details).

\vspace{8pt}

\noindent{\bf Fig. 4:\ }
$P^{*}_{[e,e]}$ for $\Lambda$ hyperons, as a function of $z$,
for different combinations of the beam and target polarizations, as
shown in the plot legend. The kinematical setup is typical of
HERMES experiment at DESY (see Table 1 for details).
Results are given for scenarios 2 (heavy lines) and 3 (thin lines)
of the polarized $\Lambda$ FF of Ref.~\cite{vog}. Results
with scenario 1 are almost negligible. Unpolarized $(\Lambda+\bar\Lambda)$
FF are from Ref.~\cite{vog}; unpolarized GRV \cite{unpdf}
and polarized GRSV-ST \cite{poldf} partonic distributions have been used 
for the proton target.
Since $\langle{x}\rangle$ is large for HERMES kinematical
configuration, we expect, at large $z$,
$T=d\sigma^{\bar\Lambda}/d\sigma^\Lambda \ll 1$ and, as a consequence,
$P_{[e,e]}(\Lambda)\simeq P^{*}_{[e,e]}(\Lambda)$, 
while for the $\bar\Lambda$ (not shown in this plot) it should result
$P_{[e,e]}(\bar\Lambda) \gg P^{*}_{[e,e]}(\bar\Lambda)$
(see text for more details).

\vspace{8pt}

\noindent{\bf Fig. 5:\ }
$P^{*}_{[\mu,\mu]}$ for $\Lambda$ hyperons, as a function of $z$,
for different combinations of the beam and target polarizations, as
shown in the plot legend. The kinematical setup is typical of
COMPASS experiment at CERN (see Table 1 for details).
Results are given for scenarios 2 (heavy lines) and 3 (thin lines)
of the polarized $\Lambda$ FF of Ref.~\cite{vog}. Results
with scenario 1 are almost negligible. Unpolarized $(\Lambda+\bar\Lambda)$
FF are from Ref.~\cite{vog}; unpolarized GRV \cite{unpdf}
and polarized GRSV-ST \cite{poldf}
partonic distributions have been used for the proton target.
Since $\langle{x}\rangle$ is relatively large for COMPASS
kinematical configuration, we expect, at large $z$, 
$T=d\sigma^{\bar\Lambda}/d\sigma^\Lambda \ll 1$ and, as a consequence,
$P_{[\mu,\mu]}(\Lambda)\simeq
P^{*}_{[\mu,\mu]}(\Lambda)$, while for the $\bar\Lambda$ (not shown
in this plot) it should result
$P_{[\mu,\mu]}(\bar\Lambda) \gg P^{*}_{[\mu,\mu]}(\bar\Lambda)$
(see text for more details).

\vspace{8pt}
\noindent{\bf Fig. 6:\ }
$P^{*}_{[\mu,\mu]}$ for $\Lambda$ hyperons, as a function of $z$,
for different combinations of the beam and target polarizations, as
shown in the plot legend. The kinematical setup is typical of
E665 experiment at SLAC (see Table 1 for details).
Results are given for scenarios 2 (heavy lines) and 3 (thin lines)
of the polarized $\Lambda$ FF of Ref.~\cite{vog}. Results
with scenario 1 are almost negligible. Unpolarized $(\Lambda+\bar\Lambda)$
FF are from Ref.~\cite{vog}; unpolarized GRV \cite{unpdf}
and polarized GRSV-ST \cite{poldf}
partonic distributions have been used for the proton target.
Since $\langle{x}\rangle$ is relatively small for E665 kinematical
configuration, we expect
$T=d\sigma^{\bar\Lambda}/d\sigma^\Lambda \simeq 1$ and, as a consequence,
$P_{[\mu,\mu]}(\Lambda,\bar\Lambda)\simeq
2\,P^{*}_{[\mu,\mu]}(\Lambda,\bar\Lambda)$
(see text for more details).

\vspace{8pt}

\noindent{\bf Fig. 7:\ }
$P^{*}_{[e,e]}$ for $\Lambda$ hyperons, as a function of $z$,
for different combinations of the positron beam and target polarizations, as
shown in the plot legend. The kinematical setup is typical of
HERA experiments at DESY (see Table 1 for details).
Results are given for scenarios 2 (heavy lines) and 3 (thin lines)
of the polarized $\Lambda$ FF of Ref.~\cite{vog}. Results
with scenario 1 are almost negligible. Unpolarized $(\Lambda+\bar\Lambda)$
FF are also from Ref.~\cite{vog}; unpolarized GRV \cite{unpdf}
and polarized GRSV-ST \cite{poldf}
partonic distributions have been used for the proton target.
Since $\langle{x}\rangle$ is small for HERA kinematical
configurations, we expect
$T=d\sigma^{\bar\Lambda}/d\sigma^\Lambda \simeq 1$ and, as a consequence,
$P_{[e,e]}(\Lambda,\bar\Lambda)\simeq
2\,P^{*}_{[e,e]}(\Lambda,\bar\Lambda)$
(see text for more details).

\vspace{8pt}

\noindent{\bf Fig. 8:\ }
$P^{*(+,0)}_{[e,e]}$ for $\Lambda$ hyperons, as a function of $x$,
both for positron (heavy lines) and electron (thin lines) beams.
The kinematical setup is typical of
HERA experiments at DESY (see Table 1 for details).
Results are given for all the three scenarios
of the polarized $\Lambda$ FF of Ref.~\cite{vog}.
Unpolarized $(\Lambda+\bar\Lambda)$
FF are from Ref.~\cite{vog}; unpolarized GRV \cite{unpdf}
partonic distributions have been used for the proton target.
The crossing at $x\simeq 0.1$ for the case of positron beam is
due to the interference between electromagnetic and weak contributions.
Since $T=d\sigma^{\bar\Lambda}/d\sigma^\Lambda$ is $\ll 1$ at large
$x$ and becomes comparable to unity at very low $x$, we expect, 
correspondingly, $P_{[e,e]}(\Lambda)\simeq P^{*}_{[e,e]}(\Lambda)$ and 
$P_{[e,e]}(\Lambda, \bar\Lambda) \simeq 2 \, P^*_{[e,e]}(\Lambda,\bar\Lambda)$
(see text and Fig. 12 for more details).

\vspace{8pt}

\noindent{\bf Fig. 9:\ }
$P^{*(0,0)}_{[e,e]}$ for $\Lambda$ hyperons, as a function of $z$,
for various ``high-$Q^2$'' options of the HERA kinematical
setup (see the HERA* setup in Table 1 for details) and for a positron beam:
$y>0.1$ (solid lines); $y>0.6$ (dashed lines); $Q^2 > 4000$ GeV$^2$ 
(dot-dashed lines).
Results are given for scenarios 2 (heavy lines) and 3 (thin lines)
of the polarized $\Lambda$ FF of Ref.~\cite{vog}. Results
with scenario 1 are almost negligible. Unpolarized $(\Lambda+\bar\Lambda)$
FF are from Ref.~\cite{vog}; unpolarized GRV \cite{unpdf}
partonic distributions have been used for the proton target
(see text for more details).

\vspace{8pt}

\noindent{\bf Fig. 10:\ }
$P^{*(+,0)}_{[e,e]}$ for $\Lambda$ hyperons, as a function of $z$,
both for positron (heavy lines) and electron (thin lines) beams,
and for various ``high-$Q^2$'' options of the HERA kinematical
setup (see the HERA* setup in Table 1 for details):
$y>0.1$ (solid lines);
$y>0.6$ (dashed lines).
Results are given for scenarios 2 and 3 
of the polarized $\Lambda$ FF of Ref.~\cite{vog}. Results
with scenario 1 are almost negligible. Unpolarized $(\Lambda+\bar\Lambda)$
FF are from Ref.~\cite{vog}; unpolarized GRV \cite{unpdf}
partonic distributions have been used for the proton target
(see text for more details).

\vspace{8pt}

\noindent{\bf Fig. 11:\ }
$P^{(0)}_{[\bar\nu,\mu]}$ for $\Lambda$ (solid lines) and
$\bar\Lambda$ (dashed lines) hyperons, as a function of $z$,
with a kinematical setup typical of NOMAD experiment at CERN
(see Table 1 for details).
Results are given for scenarios 2 and 3
of the polarized $\Lambda$ FF of Ref.~\cite{vog}. Results
with scenario 1 are almost negligible. Unpolarized $(\Lambda+\bar\Lambda)$
FF are from Ref.~\cite{vog}; unpolarized GRV \cite{unpdf}
partonic distributions have been used for the proton target.
Estimates for $P^{(0)}_{[\bar\nu,\mu]}$ are obtained from
Eqs.~(\ref{ptops}) by using the corresponding results for $P^{*(0)}_{[\bar\nu,\mu]}$,
shown in Fig. 2, and evaluating 
$T=d\sigma^{\bar\Lambda}/d\sigma^\Lambda$ 
with the $\Lambda$, $\bar\Lambda$ unpolarized FF of
Ref.~\cite{blt} (heavy lines) and Ref.~\cite{ind} (thin lines);
this last set has been modified by imposing $SU(3)$ symmetry. 
The spread between the two corresponding sets of curves gives a good 
indication of the uncertainty due to the evaluation of the ratio
$T$. Notice that this uncertainty is almost negligible for
large $z$, where polarizations are expected to be sizeable for
both scenarios 2 and 3.

\vspace{8pt}

\noindent{\bf Fig. 12:\ }
$P^{(+,0)}_{[e,e]}$ for $\Lambda$ hyperons, as a function of $x$,
both for positron (heavy lines) and electron (thin lines) beams.
The kinematical setup is typical of
HERA experiments at DESY (see Table 1 for details).
Results are given for all the three scenarios
of the polarized $\Lambda$ FF of Ref.~\cite{vog}.
Unpolarized $(\Lambda+\bar\Lambda)$
FF are from Ref.~\cite{vog}; unpolarized GRV \cite{unpdf}
partonic distributions have been used for the proton target.
Estimates for $P^{(+,0)}_{[e,e]}$ are obtained from
Eqs.~(\ref{ptops}) by using the corresponding results for $P^{*(+,0)}_{[e,e]}$,
shown in Fig. 8, and evaluating 
$T=d\sigma^{\bar\Lambda}/d\sigma^\Lambda$ 
with the $\Lambda$ unpolarized FF of
Ref.~\cite{blt}.
The crossing at $x\simeq 0.1$ for the case of positron beam is
due to the interference between electromagnetic and weak contributions.
Since $T=d\sigma^{\bar\Lambda}/d\sigma^\Lambda$ is $\ll 1$ at large
$x$ and becomes comparable to unity at very low $x$, we find,
correspondingly, $P_{[e,e]}(\Lambda)\simeq P^{*}_{[e,e]}(\Lambda)$ 
and $P_{[e,e]}(\Lambda,\bar\Lambda)\simeq
2\,P^{*}_{[e,e]}(\Lambda,\bar\Lambda)$
(see text and Fig. 8 for more details).


\newpage

\begin{figure}[t]
\begin{center}
\hspace*{1cm}
\mbox{~\epsfig{file=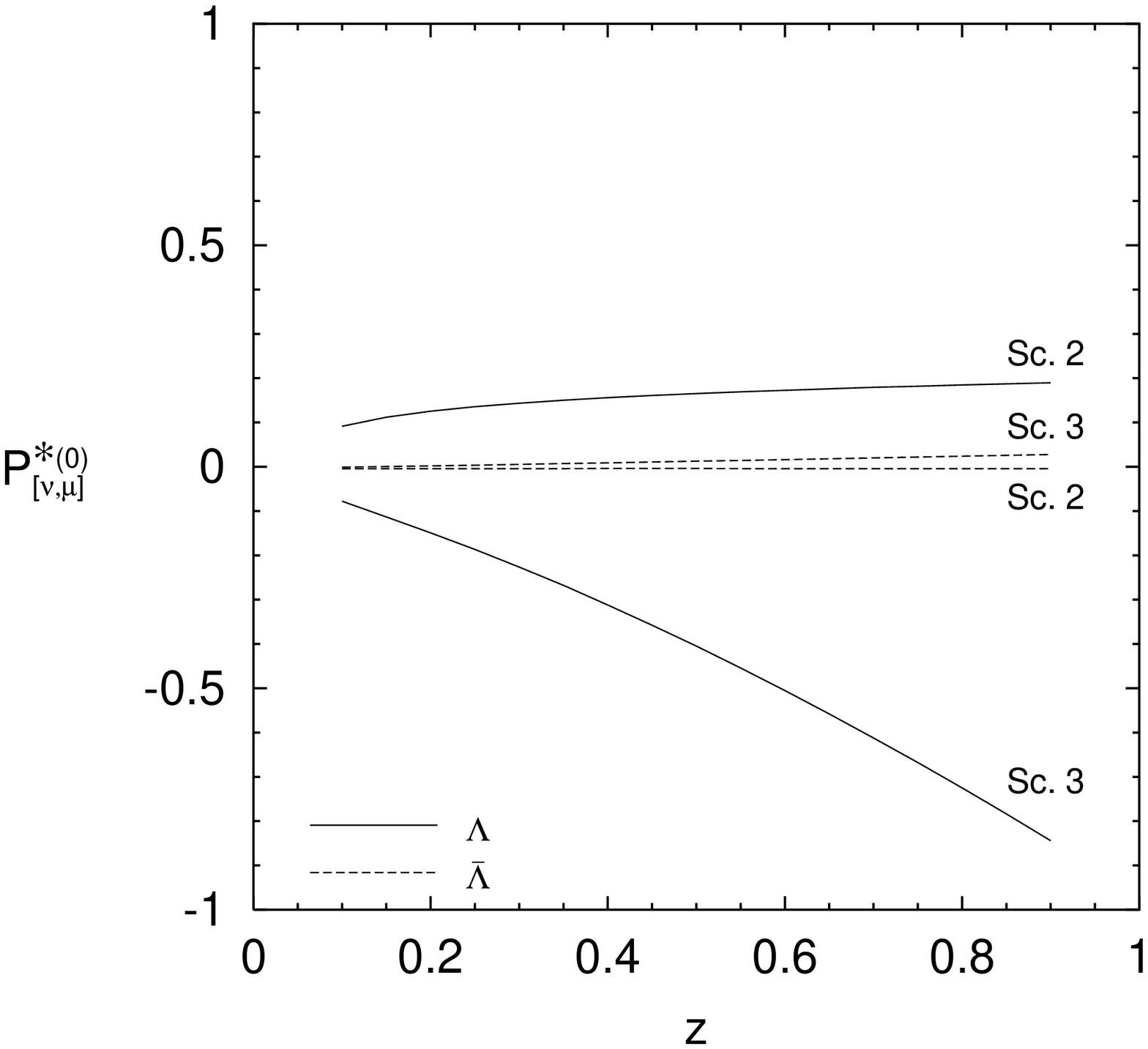,angle=-90,width=12.cm}} 
\end{center}
\caption[a1]{\label{fig1}
$P^{\ast\,(0)}_{[\nu,\mu]}$ for $\Lambda$ (solid lines) and
$\bar\Lambda$ (dashed lines) hyperons, as a function of $z$,
with a kinematical setup typical of NOMAD experiment at CERN
(see Table 1 for details).
Results are given for scenarios 2 and 3 of the polarized $\Lambda$ 
fragmentation functions (FF) of Ref.~\cite{vog}. Results
with scenario 1 are almost negligible. Unpolarized $(\Lambda+\bar\Lambda)$
FF are from Ref.~\cite{vog}; unpolarized GRV \cite{unpdf}
partonic distributions have been used for the proton target.
Since $\langle{x}\rangle$ is large for this kinematical
configuration, we expect
$T=d\sigma^{\bar\Lambda}/d\sigma^\Lambda \ll 1$ and, as a consequence,
$P^{(0)}_{[\nu,\mu]}(\Lambda)\simeq
P^{*\,(0)}_{[\nu,\mu]}(\Lambda)$, while
$P^{(0)}_{[\nu,\mu]}(\bar\Lambda) \gg
P^{*\,(0)}_{[\nu,\mu]}(\bar\Lambda)$ (see text for more details).
}
\end{figure}

\newpage

\begin{figure}[t]
\begin{center}
\hspace*{1cm}
\mbox{~\epsfig{file=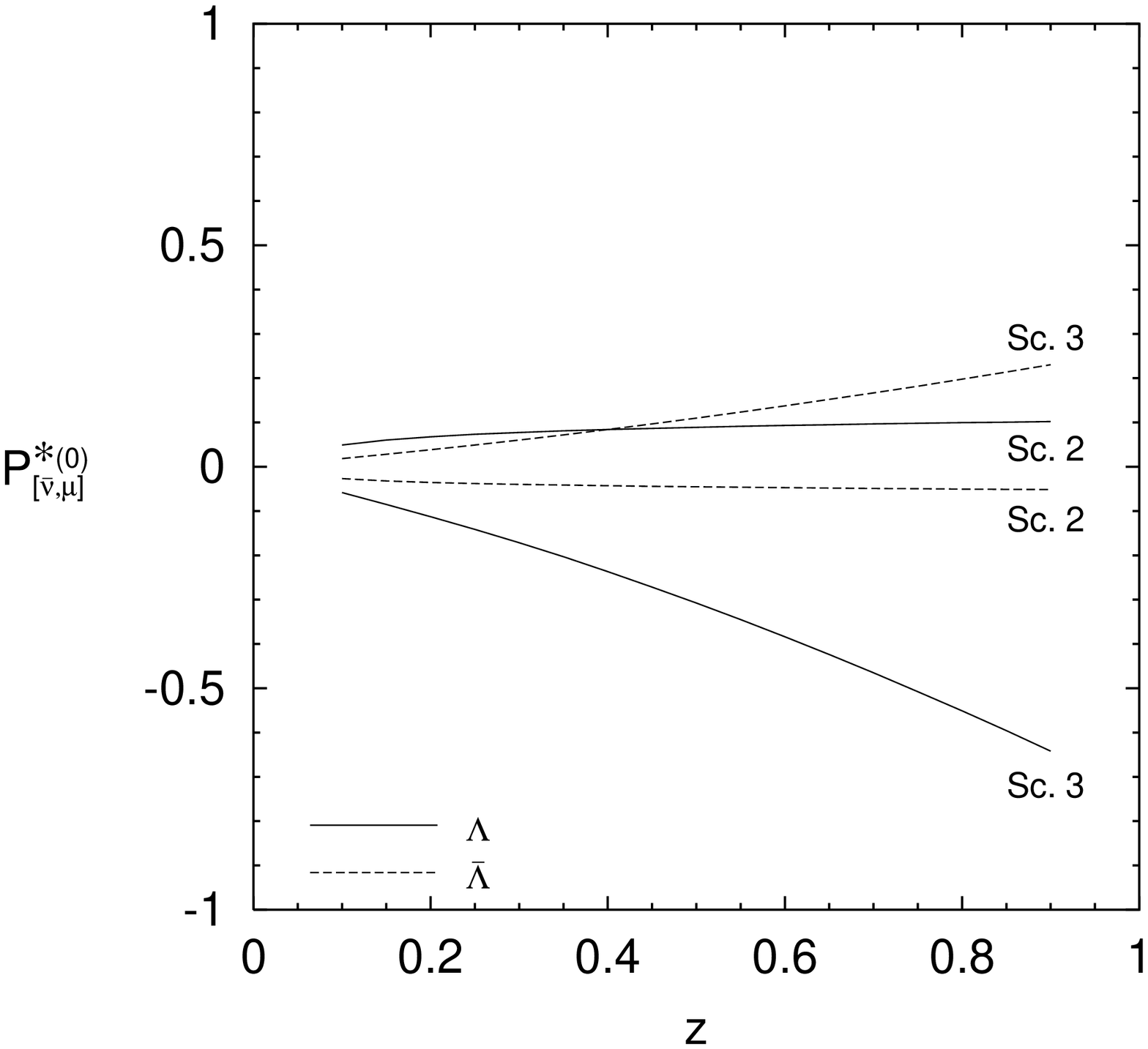,angle=-90,width=12.cm}} 
\end{center}
\caption[a2]{\label{fig2}
$P^{*\,(0)}_{[\bar\nu,\mu]}$ for $\Lambda$ (solid lines) and
$\bar\Lambda$ (dashed lines) hyperons, as a function of $z$,
with a kinematical setup typical of NOMAD experiment at CERN
(see Table 1 for details).
Results are given for scenarios 2 and 3
of the polarized $\Lambda$ FF of Ref.~\cite{vog}. Results
with scenario 1 are almost negligible. Unpolarized $(\Lambda+\bar\Lambda)$
FF are from Ref.~\cite{vog}; unpolarized GRV \cite{unpdf}
partonic distributions have been used for the proton target.
Since $\langle{x}\rangle$ is large for this kinematical
configuration, we expect
$T=d\sigma^{\bar\Lambda}/d\sigma^\Lambda \ll 1$ and, as a consequence,
$P^{(0)}_{[\bar\nu,\mu]}(\Lambda)\simeq
P^{*\,(0)}_{[\bar\nu,\mu]}(\Lambda)$, while
$P^{(0)}_{[\bar\nu,\mu]}(\bar\Lambda) \gg
P^{*\,(0)}_{[\bar\nu,\mu]}(\bar\Lambda)$
(see text and Fig. 11 for more details).
}
\end{figure}

\newpage

\begin{figure}[t]
\begin{center}
\hspace*{1cm}
\mbox{~\epsfig{file=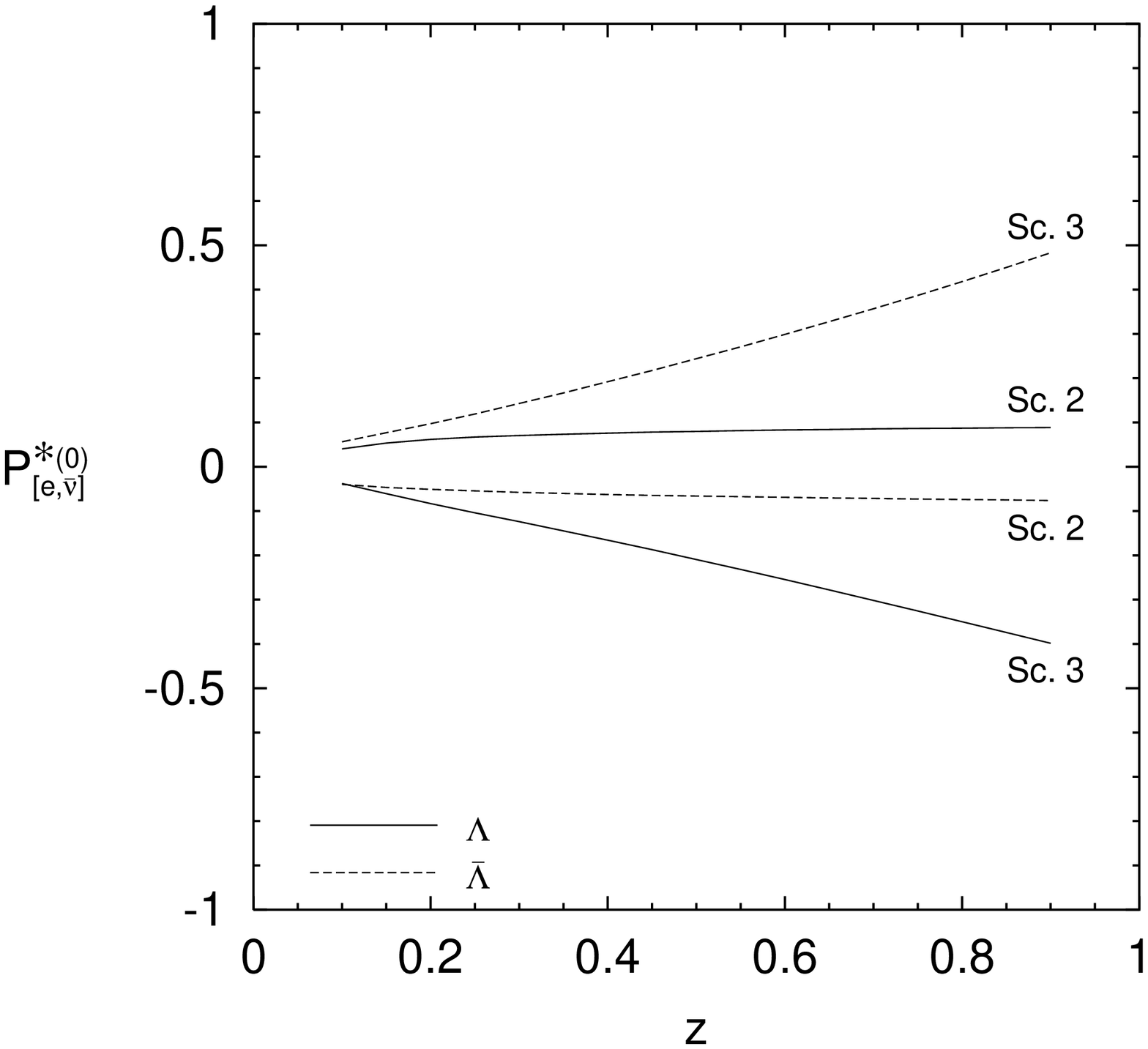,angle=-90,width=12.cm}} 
\end{center}
\caption[a3]{\label{fig3}
$P^{*\,(0)}_{[e,\bar\nu]}$ for $\Lambda$ (solid lines) and
$\bar\Lambda$ (dashed lines) hyperons, as a function of $z$,
with a kinematical setup typical of HERA experiments at DESY
(see Table 1 for details).
Results are given for scenarios 2 and 3
of the polarized $\Lambda$ FF of Ref.~\cite{vog}. Results
with scenario 1 are almost negligible. Unpolarized $(\Lambda+\bar\Lambda)$
FF are from Ref.~\cite{vog}; unpolarized GRV \cite{unpdf}
partonic distributions have been used for the proton target.
Since $\langle{x}\rangle$ is small for HERA kinematical
configurations, we expect
$T=d\sigma^{\bar\Lambda}/d\sigma^\Lambda \simeq 1$ and, as a consequence,
$P^{(0)}_{[e,\bar\nu]}(\Lambda,\bar\Lambda)\simeq
2\,P^{*\,(0)}_{[e,\bar\nu]}(\Lambda,\bar\Lambda)$
(see text for more details).
}
\end{figure}

\newpage

\begin{figure}[t]
\begin{center}
\hspace*{1cm}
\mbox{~\epsfig{file=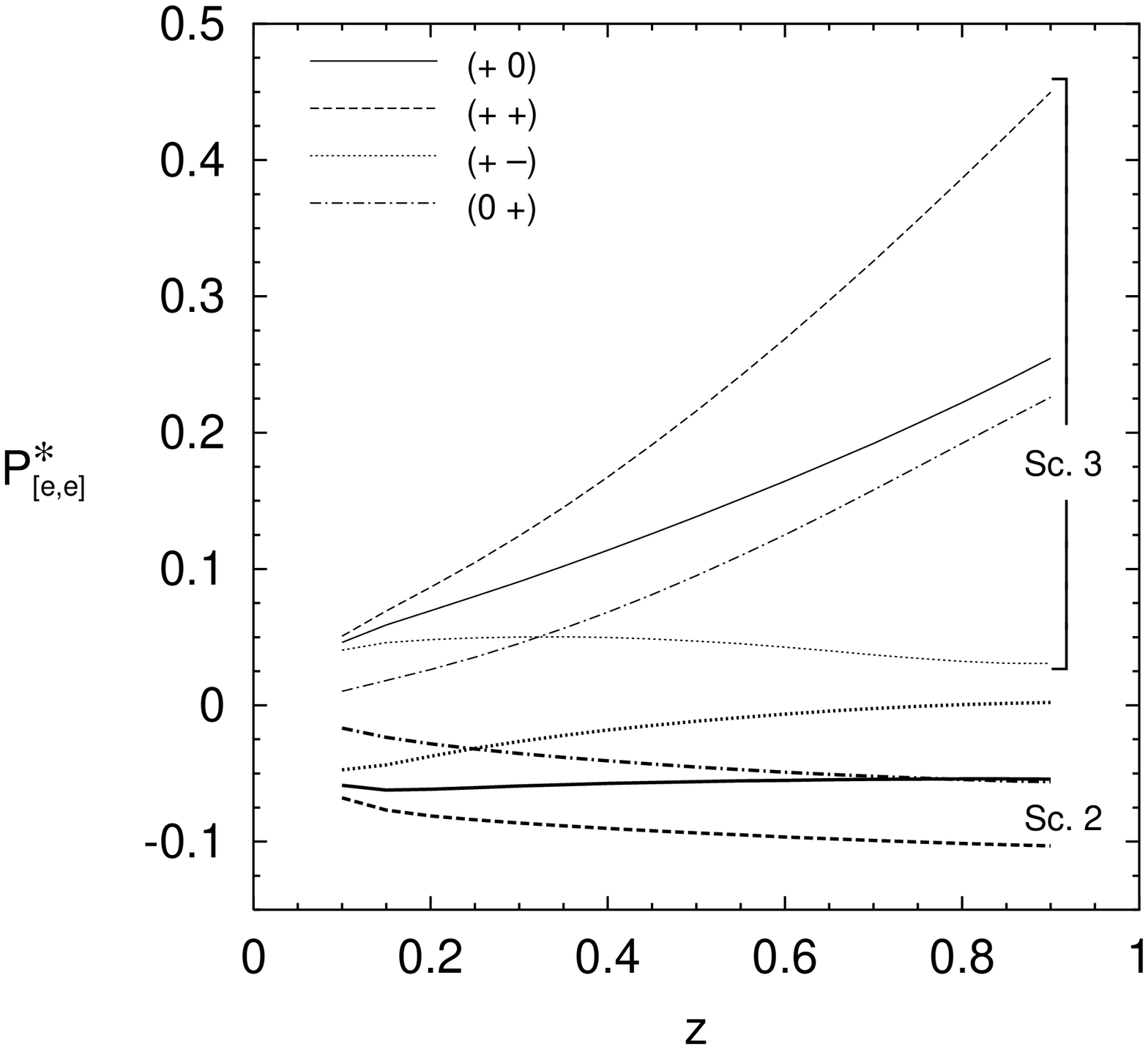,angle=-90,width=12.cm}} 
\end{center}
\caption[a4]{\label{fig4}
$P^{*}_{[e,e]}$ for $\Lambda$ hyperons, as a function of $z$,
for different combinations of the beam and target polarizations, as
shown in the plot legend. The kinematical setup is typical of
HERMES experiment at DESY (see Table 1 for details).
Results are given for scenarios 2 (heavy lines) and 3 (thin lines)
of the polarized $\Lambda$ FF of Ref.~\cite{vog}. Results
with scenario 1 are almost negligible. Unpolarized $(\Lambda+\bar\Lambda)$
FF are from Ref.~\cite{vog}; unpolarized GRV \cite{unpdf}
and polarized GRSV-ST \cite{poldf} partonic distributions have been used 
for the proton target.
Since $\langle{x}\rangle$ is large for HERMES kinematical
configuration, we expect, at large $z$,
$T=d\sigma^{\bar\Lambda}/d\sigma^\Lambda \ll 1$ and, as a consequence,
$P_{[e,e]}(\Lambda)\simeq P^{*}_{[e,e]}(\Lambda)$, 
while for the $\bar\Lambda$ (not shown in this plot) it should result
$P_{[e,e]}(\bar\Lambda) \gg P^{*}_{[e,e]}(\bar\Lambda)$
(see text for more details).
}
\end{figure}

\newpage

\begin{figure}[t]
\begin{center}
\hspace*{1cm}
\mbox{~\epsfig{file=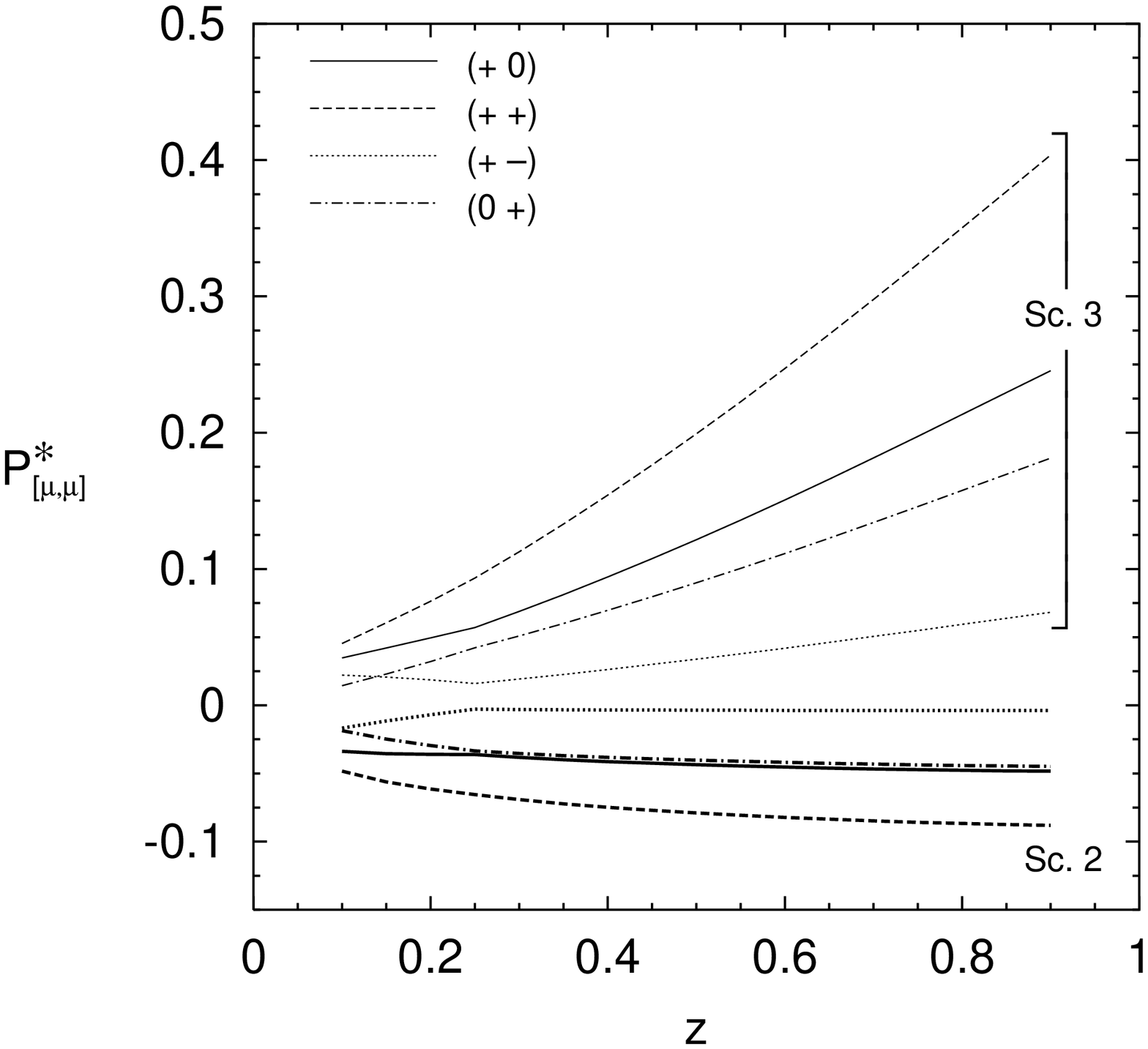,angle=-90,width=12.cm}} 
\end{center}
\caption[a5]{\label{fig5}
$P^{*}_{[\mu,\mu]}$ for $\Lambda$ hyperons, as a function of $z$,
for different combinations of the beam and target polarizations, as
shown in the plot legend. The kinematical setup is typical of
COMPASS experiment at CERN (see Table 1 for details).
Results are given for scenarios 2 (heavy lines) and 3 (thin lines)
of the polarized $\Lambda$ FF of Ref.~\cite{vog}. Results
with scenario 1 are almost negligible. Unpolarized $(\Lambda+\bar\Lambda)$
FF are from Ref.~\cite{vog}; unpolarized GRV \cite{unpdf}
and polarized GRSV-ST \cite{poldf}
partonic distributions have been used for the proton target.
Since $\langle{x}\rangle$ is relatively large for COMPASS
kinematical configuration, we expect, at large $z$, 
$T=d\sigma^{\bar\Lambda}/d\sigma^\Lambda \ll 1$ and, as a consequence,
$P_{[\mu,\mu]}(\Lambda)\simeq
P^{*}_{[\mu,\mu]}(\Lambda)$, while for the $\bar\Lambda$ (not shown
in this plot) it should result
$P_{[\mu,\mu]}(\bar\Lambda) \gg P^{*}_{[\mu,\mu]}(\bar\Lambda)$
(see text for more details).
}
\end{figure}

\newpage

\begin{figure}[t]
\begin{center}
\hspace*{1cm}
\mbox{~\epsfig{file=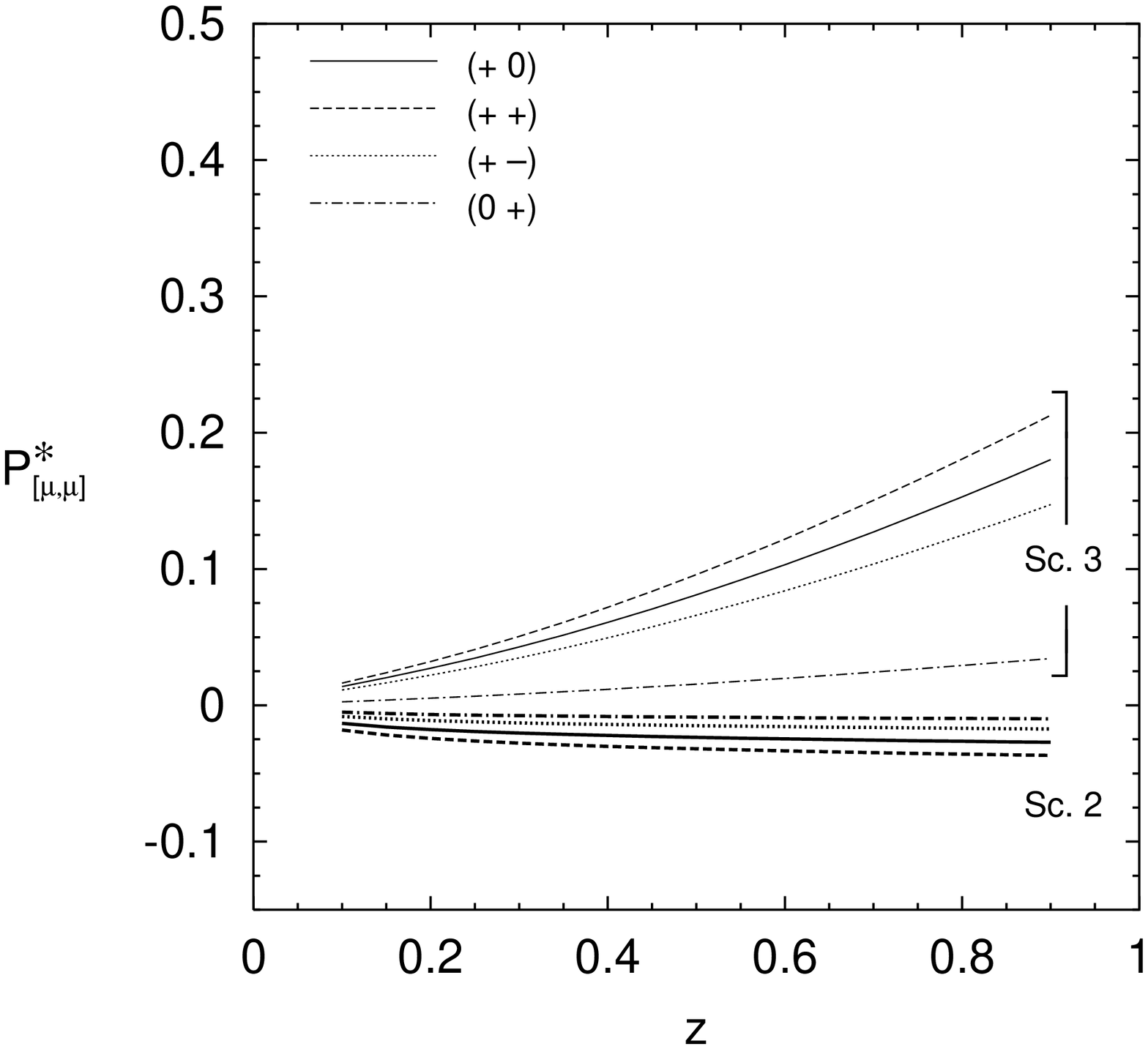,angle=-90,width=12.cm}} 
\end{center}
\caption[a6]{\label{fig6}
$P^{*}_{[\mu,\mu]}$ for $\Lambda$ hyperons, as a function of $z$,
for different combinations of the beam and target polarizations, as
shown in the plot legend. The kinematical setup is typical of
E665 experiment at SLAC (see Table 1 for details).
Results are given for scenarios 2 (heavy lines) and 3 (thin lines)
of the polarized $\Lambda$ FF of Ref.~\cite{vog}. Results
with scenario 1 are almost negligible. Unpolarized $(\Lambda+\bar\Lambda)$
FF are from Ref.~\cite{vog}; unpolarized GRV \cite{unpdf}
and polarized GRSV-ST \cite{poldf}
partonic distributions have been used for the proton target.
Since $\langle{x}\rangle$ is relatively small for E665 kinematical
configuration, we expect
$T=d\sigma^{\bar\Lambda}/d\sigma^\Lambda \simeq 1$ and, as a consequence,
$P_{[\mu,\mu]}(\Lambda,\bar\Lambda)\simeq
2\,P^{*}_{[\mu,\mu]}(\Lambda,\bar\Lambda)$
(see text for more details).
}
\end{figure}

\newpage

\begin{figure}[t]
\begin{center}
\hspace*{1cm}
\mbox{~\epsfig{file=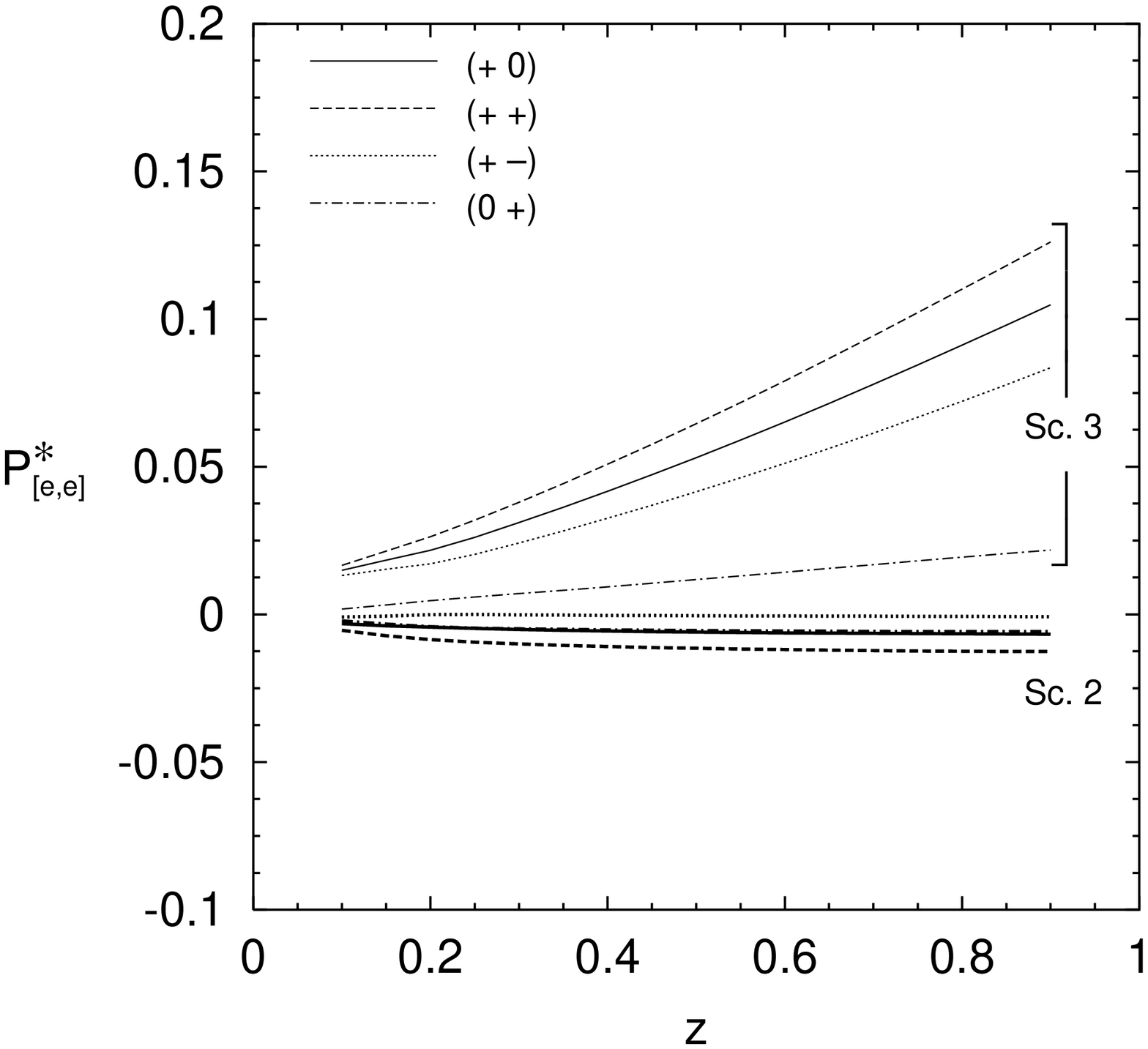,angle=-90,width=12.cm}} 
\end{center}
\caption[a7]{\label{fig7}
$P^{*}_{[e,e]}$ for $\Lambda$ hyperons, as a function of $z$,
for different combinations of the positron beam and target polarizations, as
shown in the plot legend. The kinematical setup is typical of
HERA experiments at DESY (see Table 1 for details).
Results are given for scenarios 2 (heavy lines) and 3 (thin lines)
of the polarized $\Lambda$ FF of Ref.~\cite{vog}. Results
with scenario 1 are almost negligible. Unpolarized $(\Lambda+\bar\Lambda)$
FF are also from Ref.~\cite{vog}; unpolarized GRV \cite{unpdf}
and polarized GRSV-ST \cite{poldf}
partonic distributions have been used for the proton target.
Since $\langle{x}\rangle$ is small for HERA kinematical
configurations, we expect
$T=d\sigma^{\bar\Lambda}/d\sigma^\Lambda \simeq 1$ and, as a consequence,
$P_{[e,e]}(\Lambda,\bar\Lambda)\simeq
2\,P^{*}_{[e,e]}(\Lambda,\bar\Lambda)$
(see text for more details).
}
\end{figure}

\newpage

\begin{figure}[t]
\begin{center}
\hspace*{1cm}
\mbox{~\epsfig{file=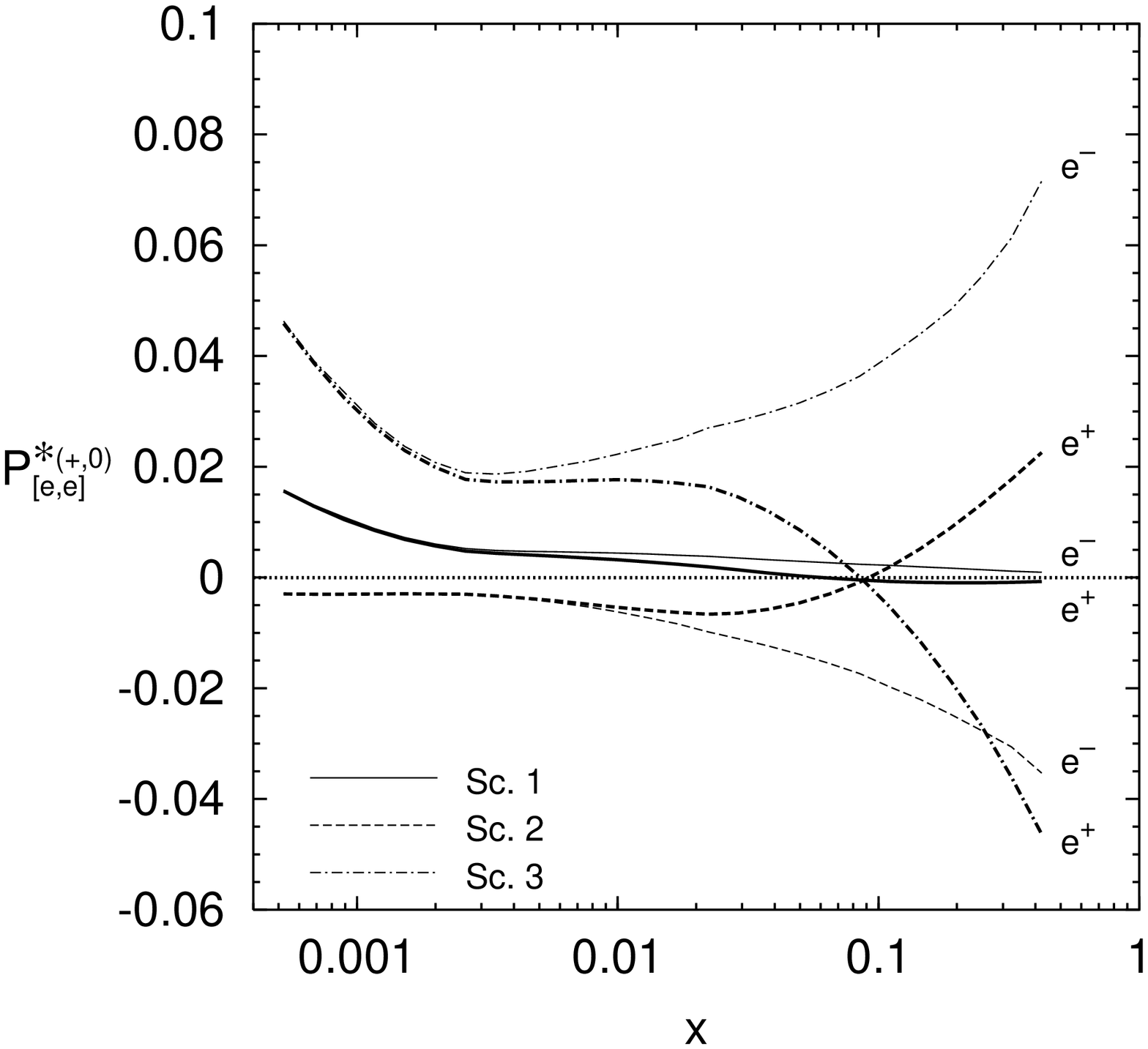,angle=-90,width=12.cm}}
\end{center}
\caption[a8]{\label{fig8}
$P^{*(+,0)}_{[e,e]}$ for $\Lambda$ hyperons, as a function of $x$,
both for positron (heavy lines) and electron (thin lines) beams.
The kinematical setup is typical of
HERA experiments at DESY (see Table 1 for details).
Results are given for all the three scenarios
of the polarized $\Lambda$ FF of Ref.~\cite{vog}.
Unpolarized $(\Lambda+\bar\Lambda)$
FF are from Ref.~\cite{vog}; unpolarized GRV \cite{unpdf}
partonic distributions have been used for the proton target.
The crossing at $x\simeq 0.1$ for the case of positron beam is
due to the interference between electromagnetic and weak contributions.
Since $T=d\sigma^{\bar\Lambda}/d\sigma^\Lambda$ is $\ll 1$ at large
$x$ and becomes comparable to unity at very low $x$, we expect, 
correspondingly, $P_{[e,e]}(\Lambda)\simeq P^{*}_{[e,e]}(\Lambda)$ and 
$P_{[e,e]}(\Lambda, \bar\Lambda) \simeq 2 \, P_{[e,e]}(\Lambda,\bar\Lambda)$
(see text and Fig. 12 for more details).
}
\end{figure}

\newpage

\begin{figure}[t]
\begin{center}
\hspace*{1cm}
\mbox{~\epsfig{file=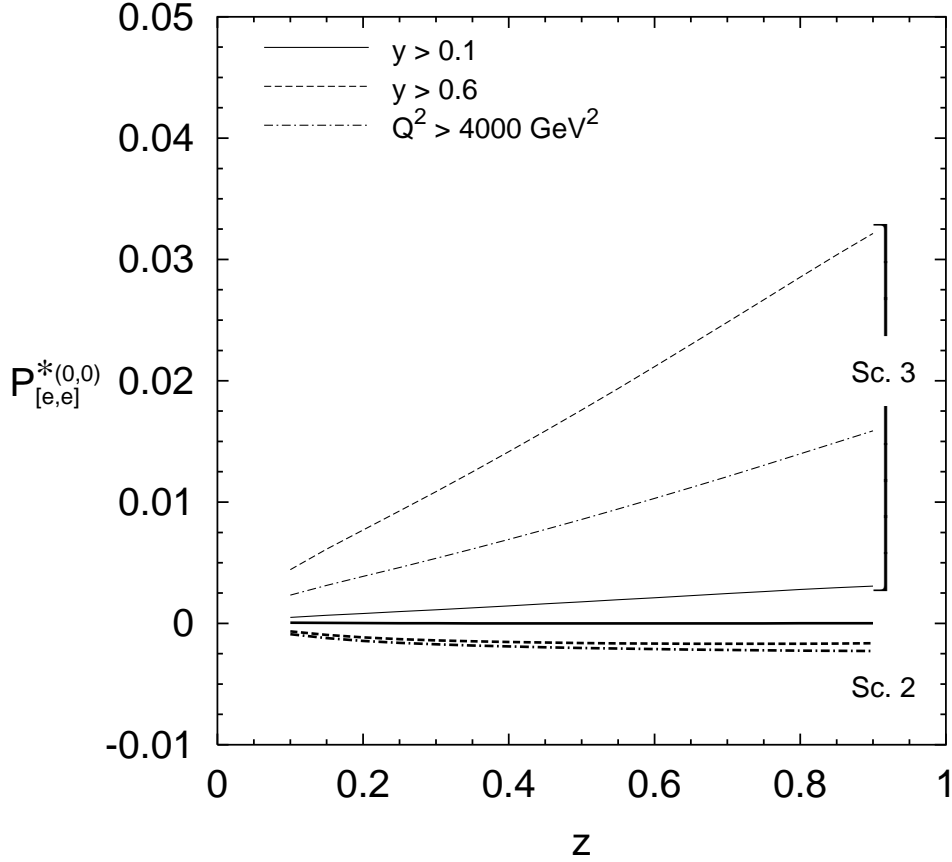,angle=-90,width=12.cm}} 
\end{center}
\caption[a9]{\label{fig9}
$P^{*(0,0)}_{[e,e]}$ for $\Lambda$ hyperons, as a function of $z$,
for various ``high-$Q^2$'' options of the HERA kinematical
setup (see the HERA* setup in Table 1 for details) and for a positron beam:
$y>0.1$ (solid lines); 
$y>0.6$ (dashed lines); $Q^2 > 4000$ GeV$^2$ (dot-dashed lines).
Results are given for scenarios 2 (heavy lines) and 3 (thin lines)
of the polarized $\Lambda$ FF of Ref.~\cite{vog}. Results
with scenario 1 are almost negligible. Unpolarized $(\Lambda+\bar\Lambda)$
FF are from Ref.~\cite{vog}; unpolarized GRV \cite{unpdf}
partonic distributions have been used for the proton target
(see text for more details).
}
\end{figure}

\newpage

\begin{figure}[t]
\begin{center}
\hspace*{1cm}
\mbox{~\epsfig{file=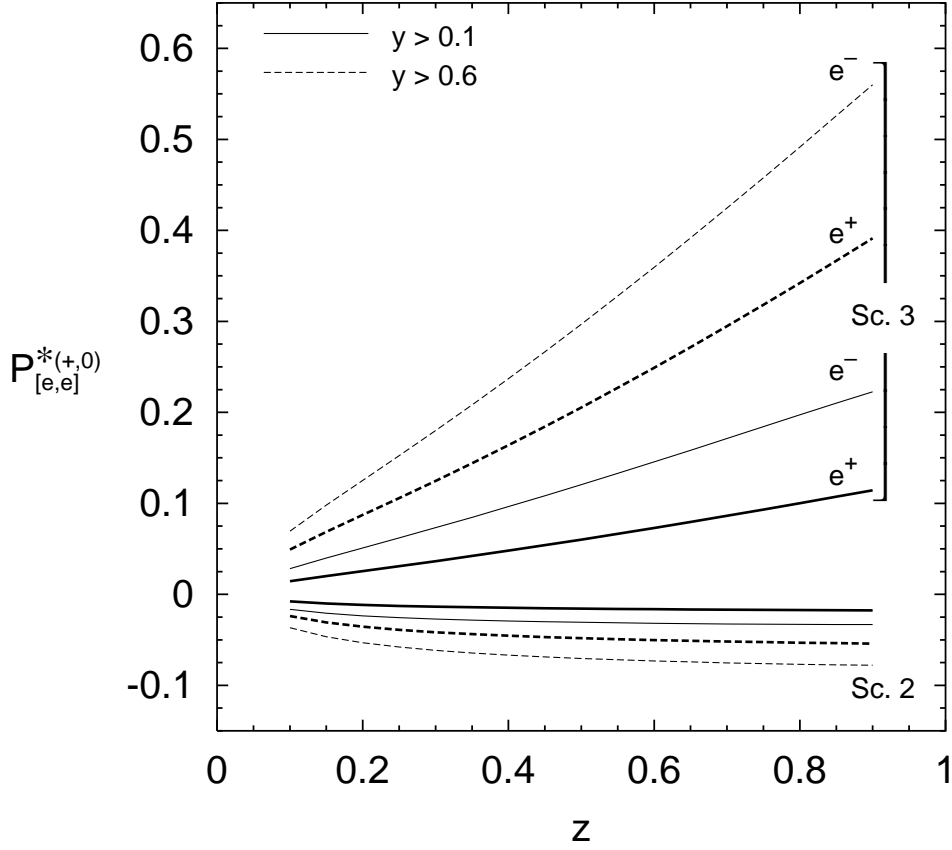,angle=-90,width=12.cm}}
\end{center}
\caption[a10]{\label{fig10}
$P^{*(+,0)}_{[e,e]}$ for $\Lambda$ hyperons, as a function of $z$,
both for positron (heavy lines) and electron (thin lines) beams,
and for various ``high-$Q^2$'' options of the HERA kinematical
setup (see the HERA* setup in Table 1 for details):
$y>0.1$ (solid lines);
$y>0.6$ (dashed lines).
Results are given for scenarios 2 and 3 
of the polarized $\Lambda$ FF of Ref.~\cite{vog}. Results
with scenario 1 are almost negligible. Unpolarized $(\Lambda+\bar\Lambda)$
FF are from Ref.~\cite{vog}; unpolarized GRV \cite{unpdf}
partonic distributions have been used for the proton target
(see text for more details).
}
\end{figure}

\newpage

\begin{figure}[t]
\begin{center}
\hspace*{1cm}
\mbox{~\epsfig{file=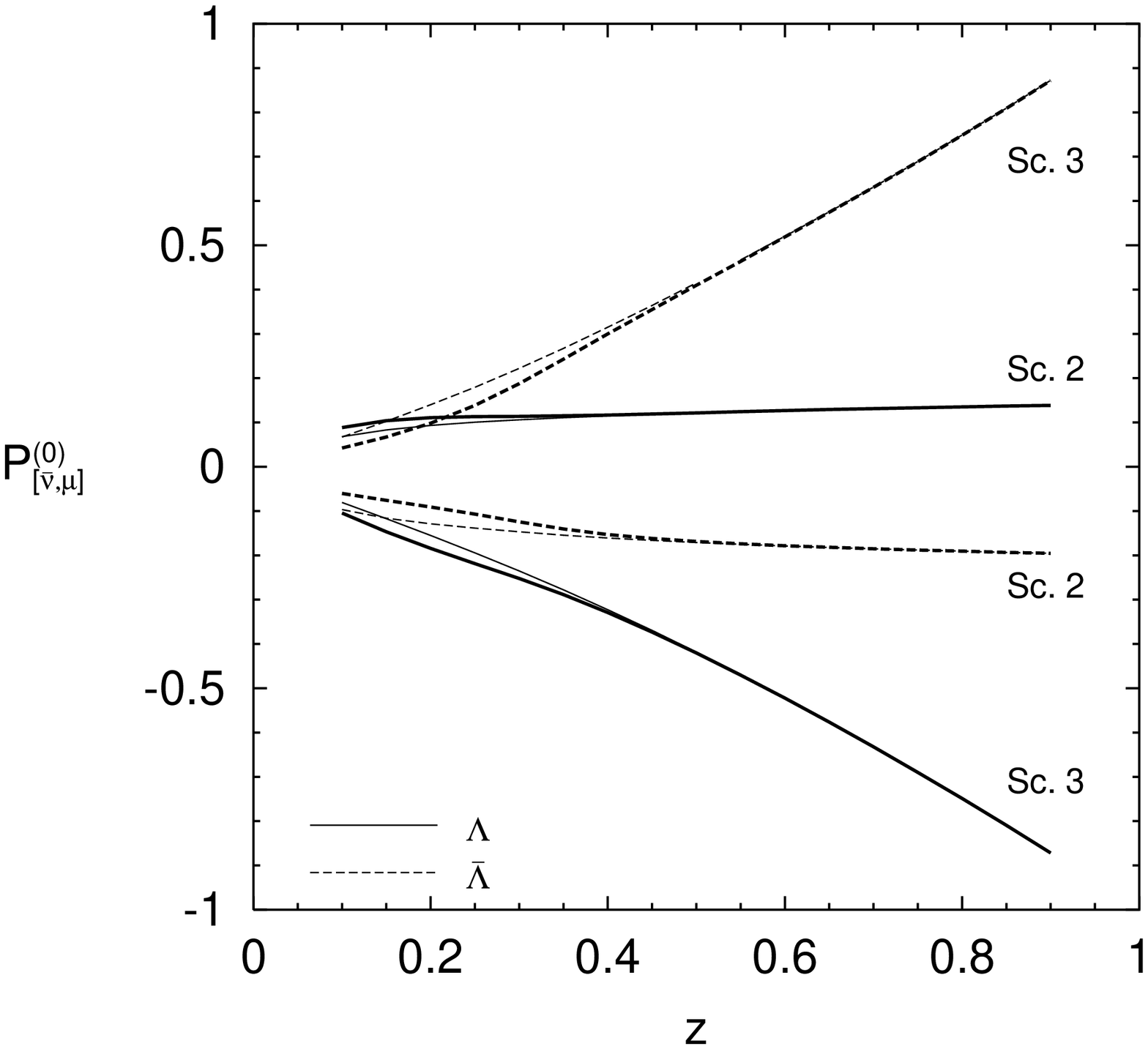,angle=-90,width=12.cm}}
\end{center}
\caption[a11]{\label{fig11}
$P^{(0)}_{[\bar\nu,\mu]}$ for $\Lambda$ (solid lines) and
$\bar\Lambda$ (dashed lines) hyperons, as a function of $z$,
with a kinematical setup typical of NOMAD experiment at CERN
(see Table 1 for details).
Results are given for scenarios 2 and 3
of the polarized $\Lambda$ FF of Ref.~\cite{vog}. Results
with scenario 1 are almost negligible. Unpolarized $(\Lambda+\bar\Lambda)$
FF are from Ref.~\cite{vog}; unpolarized GRV \cite{unpdf}
partonic distributions have been used for the proton target.
Estimates for $P^{(0)}_{[\bar\nu,\mu]}$ are obtained from
Eqs.~(\protect\ref{ptops}) by using the
corresponding results for $P^{*(0)}_{[\bar\nu,\mu]}$,
shown in Fig. 2, and evaluating 
$T=d\sigma^{\bar\Lambda}/d\sigma^\Lambda$ 
with the $\Lambda$, $\bar\Lambda$ unpolarized FF of
Ref.~\cite{blt} (heavy lines) and Ref.~\cite{ind} (thin lines);
this last set has been modified by imposing $SU(3)$ symmetry. 
The spread between the two corresponding sets of curves gives a good 
indication of the uncertainty due to the evaluation of the ratio
$T$. Notice that this uncertainty is almost negligible for
large $z$, where polarizations are expected to be sizeable for
both scenarios 2 and 3.
}
\end{figure}

\newpage

\begin{figure}[t]
\begin{center}
\hspace*{1cm}
\mbox{~\epsfig{file=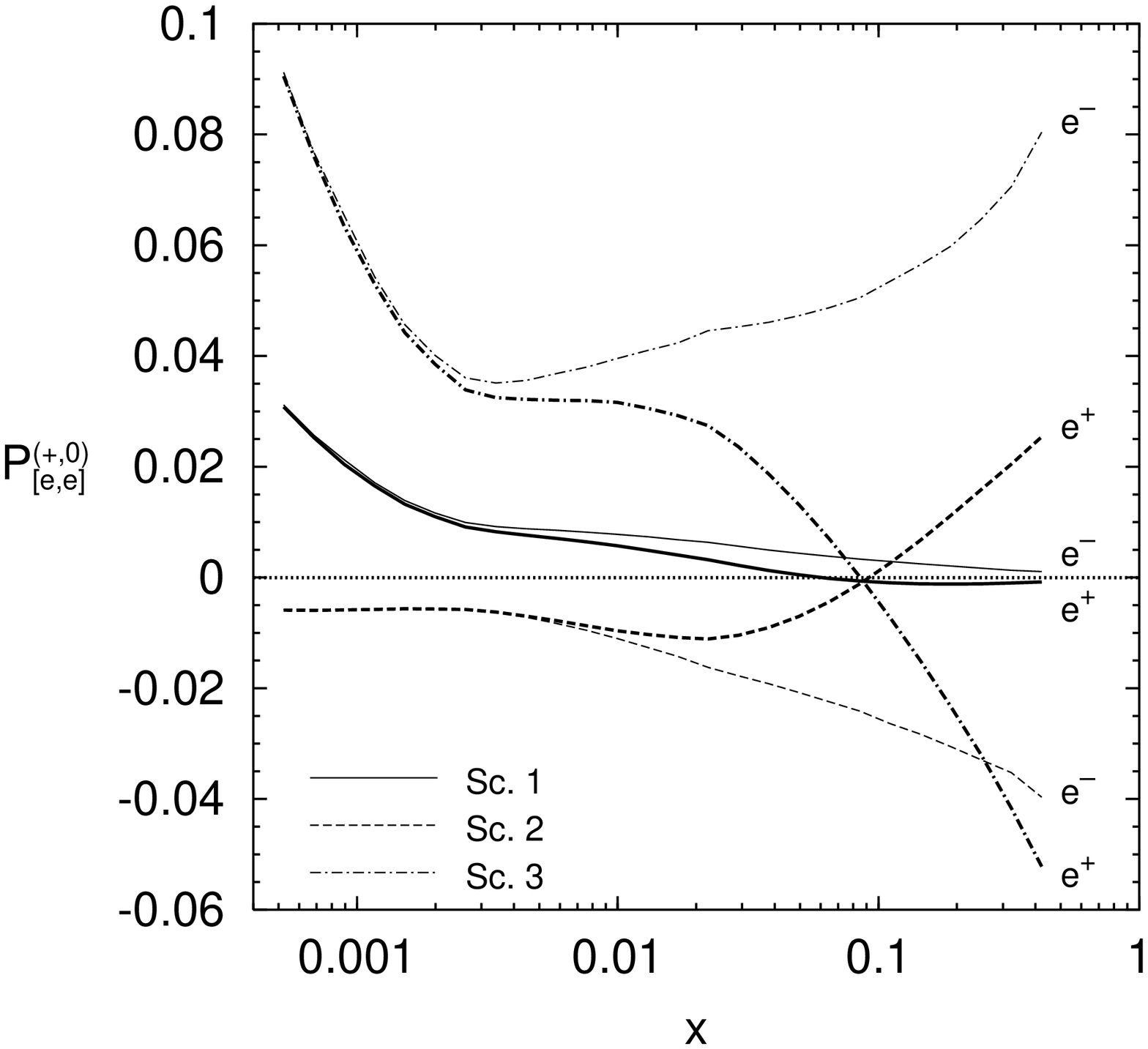,angle=-90,width=12.cm}}
\end{center}
\caption[a12]{\label{fig12}
$P^{(+,0)}_{[e,e]}$ for $\Lambda$ hyperons, as a function of $x$,
both for positron (heavy lines) and electron (thin lines) beams.
The kinematical setup is typical of
HERA experiments at DESY (see Table 1 for details).
Results are given for all the three scenarios
of the polarized $\Lambda$ FF of Ref.~\cite{vog}.
Unpolarized $(\Lambda+\bar\Lambda)$
FF are from Ref.~\cite{vog}; unpolarized GRV \cite{unpdf}
partonic distributions have been used for the proton target.
Estimates for $P^{(+,0)}_{[e,e]}$ are obtained from
Eqs.~(\protect\ref{ptops}) by using the corresponding
results for $P^{*(+,0)}_{[e,e]}$,
shown in Fig. 8, and evaluating 
$T=d\sigma^{\bar\Lambda}/d\sigma^\Lambda$ 
with the $\Lambda$ unpolarized FF of
Ref.~\cite{blt}.
The crossing at $x\simeq 0.1$ for the case of positron beam is
due to the interference between electromagnetic and weak contributions.
Since $T=d\sigma^{\bar\Lambda}/d\sigma^\Lambda$ is $\ll 1$ at large
$x$ and becomes comparable to unity at very low $x$, we find,
correspondingly, $P_{[e,e]}(\Lambda)\simeq P^{*}_{[e,e]}(\Lambda)$ 
and $P_{[e,e]}(\Lambda,\bar\Lambda)\simeq
2\,P^{*}_{[e,e]}(\Lambda,\bar\Lambda)$
(see text and Fig. 8 for more details).
}
\end{figure}

\newpage

\vskip 36pt
\baselineskip=6pt
\begin{thebibliography}{99}
\small
\vspace{6pt}
\setlength{\parskip}{6pt}
\bibitem{noi1}
\vskip-8pt
M. Anselmino, M. Boglione, J. Hansson and F. Murgia, \PR{D54} (1996) 828
\bibitem{jaf}
\vskip-8pt
R.L. Jaffe, \PR{D54} (1996) 6581
\bibitem{ekk}
\vskip-8pt
J. Ellis, D. Kharzeev and A. Kotzinian, \ZP{C69} (1996) 467
\bibitem{vog}
\vskip-8pt
D. de Florian, M. Stratmann and W. Vogelsang, \PR{D57} (1998) 5811 
\bibitem{kbv}
\vskip-8pt
A. Kotzinian, A. Bravar and D. von Harrach,
{\it Eur. Phys. J.} {\bf C2} (1998) 329
\bibitem{kot}
\vskip-8pt
A. Kotzinian, talk at the VII Workshop on {\it High Energy
Spin Physics} (SPIN-97), 7-12 July 1997, Dubna, Russia;
e-Print Archive: hep-ph/9709259
\bibitem{bel}
\vskip-8pt
S.L. Belostotski, \NP{B79} (Proc. Suppl.) (1999) 526
\bibitem{bjk}
\vskip-8pt
D. Boer, R. Jakob and P.J. Mulders, \NP{B564} (2000) 471 
\bibitem{blt}
\vskip-8pt
C. Boros, J.T. Londergan and A.W. Thomas, \PR{D61} (2000) 014007
\bibitem{al}
\vskip-8pt
D. Ashery and H.J. Lipkin, \PL{B469} (1999) 263
\bibitem{mssy}
\vskip-8pt
B-Q. Ma, I. Schmidt, J. Soffer and J-Y. Yang, {\it Eur. Phys. J.}
 {\bf C16} (2000) 657; {\it Phys. Rev.} {\bf D62} (2000) 114009 
\bibitem{noi2}
\vskip-8pt
M. Anselmino, M. Boglione and F. Murgia, \PL{B481} (2000) 253 
\bibitem{nom}
\vskip-8pt
NOMAD Collaboration, P. Astier {\it et al.}, \NP{B588} (2000) 3; 
e-Print Archive: hep-ex/0103047
\bibitem{unpdf}
\vskip-8pt
M. Gl\"uck, E. Reya and A. Vogt, \ZP{C67} (1995) 433
\bibitem{poldf}
\vskip-8pt
M. Gl\"uck, E. Reya, M. Stratmann and W. Vogelsang, \PR{D53} (1996) 4775
\bibitem{ind}
\vskip-8pt
D. Indumathi, H.S. Mani and A. Rastogi, \PR{D58} (1998) 094014
\bibitem{rep}
\vskip-8pt
M.L. Mangano {\it et al.}, CERN report CERN-TH/2001-131,
 e-Print Archive: hep-ex/0105155 
\bibitem{spfl}
\vskip-8pt
M.~Anselmino, M.~Boglione, U.~D'Alesio, E.~Leader and F.~Murgia, 
\PL{B509} (2001) 246
\bibitem{nau}
\vskip-8pt
D.V. Naumov, private communication
\end{thebibliography}
\end{document}